\documentclass[useAMS,usenatbib]{mn2e}
\usepackage{natbib}
\usepackage{amsmath,amsfonts}
\usepackage{epsfig}
\usepackage{mathptmx}

\title[AMI-LA observations of cores]
{AMI Large Array radio continuum observations of \emph{Spitzer} c2d small clouds and cores\thanks{We request that any reference to this paper cites ``AMI Consortium: Scaife et~al. 2010"}}

\label{firstpage}
\author[Scaife et~al.]
{AMI Consortium: Anna M. M. Scaife$^1$$\thanks{E-mail: 
ascaife@cp.dias.ie}$,
 Emily I. Curtis$^{2,3}$,
 Matthew Davies$^2$,
\newauthor
 Thomas M. O. Franzen$^2$,
 Keith J. B. Grainge$^{2,3}$,
 Michael P. Hobson$^2$,
 Natasha Hurley-Walker$^2$,
\newauthor
 Anthony N. Lasenby$^{2,3}$,
 Malak Olamaie$^2$,
 Guy G. Pooley$^2$,
 Carmen Rodr{\'i}guez-Gonz{\'a}lvez$^2$,
\newauthor
 Richard D. E. Saunders$^{2,3}$,
 Michel Schammel$^2$,
 Paul F. Scott$^2$,
 Timothy Shimwell$^2$,
\newauthor
 David Titterington$^2$,
 Elizabeth Waldram$^2$ \& 
 Jonathan T. L. Zwart$^4$.
 \vspace{0.03in}\\
$^1$ Dublin Institute for Advanced Studies, 31 Fitzwilliam Place,
Dublin 2, Ireland\\
$^2$ Astrophysics Group, Cavendish Laboratory, J J Thomson Avenue,
Cambridge CB3 0HE\\
$^3$ Kavli Institute for Cosmology Cambridge, Madingley Road,
Cambridge, CB3 0HA\\
$^4$ Columbia Astrophysics Laboratory, Columbia University, 550 West 120th
Street, New York 10027, USA\\
}

\date{Accepted ---; received ---; in original form \today}
\pagerange{\pageref{firstpage}--\pageref{lastpage}}
\pubyear{2005}

\begin{document}
\maketitle

\begin{abstract}
\noindent
We perform deep 1.8\,cm radio continuum imaging towards thirteen protostellar regions selected from the \emph{Spitzer} c2d small clouds and cores programme at high resolution ($25''$) in order to detect and quantify the cm--wave emission from deeply embedded young protostars. Within these regions we detect fifteen compact radio sources which we identify as radio protostars including two probable new detections. The sample is in general of low bolometric luminosity and contains several of the newly detected VeLLO sources. We determine the 1.8\,cm radio luminosity to bolometric luminosity correlation, $L_{\rm{rad}}-L_{\rm{bol}}$, for the sample and discuss the nature of the radio emission in terms of the available sources of ionized gas. We also investigate the $L_{\rm{rad}}-L_{\rm{IR}}$ correlation and suggest that radio flux density may be used as a proxy for the internal luminosity of low luminosity protostars.
\end{abstract}

\begin{keywords}
Radiation mechanisms:general -- ISM:general -- ISM:clouds -- stars:formation
\end{keywords}

\section{Introduction}

The radio spectrum of dark clouds has recently come under scrutiny following the proposal that their centimetre-wave emission is due to spinning dust (Casassus et~al. 2006; AMI Consortium: Scaife et~al. 2009a, b; 2010). These observational studies have largely concentrated on the arcminute scale emission, neglecting the known small scale radio emission detected from a significant fraction of protostars (Anglada 1995; Rodr{\'i}guez et~al. 1989). As these investigations become more detailed it is necessary to quantify correctly this emission in order to avoid confusing it with that arising from spinning dust when observing at resolutions where the two may not be easily distinguishable. Most radio protostar searches and surveys (e.g. Anglada 1995; Stamatellos et~al. 2009; Andr{\'e} et~al. 1987) are conducted at 3.6 and 6\,cm wavelengths, however the spectrum of emission due to spinning dust is thought to peak at shorter wavelengths of 1--2\,cm (Drain \& Lazarian 1998) and this is the wavelength range where most spinning dust observations are made. Although a few observations towards specific sources have been made at 2\,cm (Rodr{\'i}guez \& Canto 1983; Bieging, Cohen \& Schwartz 1984; Pravdo 1985; Rodr{\'i}guez \& Reipurth 1997; Andr{\'e}, Motte \& Bacmann 1999) their number is still limited, particularly in the case of low luminosity protostars and Class O objects.

Longer wavelength radio emission is capable of escaping from the high column density envelopes which surround deeply embedded young protostars, making it an appropriate mechanism for searching for such objects. Very high sensitivity observations are required as the thermal (vibrational) dust spectrum falls off steeply in intensity at longer wavelengths. However, young and low luminosity protostars are also known to produce additional radio emission which can make them more easily detectable. In general this radio emission has been observed to possess a rising spectrum, indicating that it occurs as a consequence of free--free radiation from ionized gas. The spectral index of such emission (defined as $\alpha$, where $S_{\nu}\propto \nu^{\alpha}$) can lie anywhere in the range $-0.1\leq \alpha \leq 2$. An index of $\alpha=-0.1$ indicates optically thin free-free emission, whereas $\alpha=2$ indicates optically thick; values intermediate to these limits imply a medium which is partially optically thick (Reynolds 1986). In the case of a spherically symmetric shell of decreasing density with $n_{\rm{e}}\propto r^{-2}$, $\alpha \simeq 0.6$.  A number of mechanisms have been proposed to produce free--free emission in the immediate vicinity of stellar objects: where a high enough ionizing flux is present, generally in later type T Tauri stars, photoionization can support an embedded {\sc Hii} region (Churchwell 1990); a fully ionized stellar wind, again associated more often with later type stars, would produce an easily observable radio signal (Panagia \& Felli 1975; Curiel et~al. 1989), as could a partially ionized, collimated outflow (Reynolds 1986); the accretion shock on the surface of protostellar discs may heat and ionize infalling gas (Winkler \& Newman 1980; Cassen \& Moosman 1981); the molecular outflows from young, low mass protostars and the neutral winds thought to initiate them can generate free--free emission through shock ionization as they impact on the surrounding envelope (Curiel et~al. 1987; 1989; Rodr{\'i}guez \& Reipurth 1996). In the case of low luminosity protostars it is the last of these mechanisms which is considered to be most plausible and indeed, where measured, the outflow force of most protostellar jets has been found to be energetically viable to explain the observed cm-wave radio emission (Anglada 1995). This theory is also supported by very high angular resolution observations of the radio emission from protostars (Simon et~al. 1983; Anglada 1996), which find that the radio emission is elongated along the direction of the molecular outflows.

Recent measurements of the luminosity distribution of protostars from the \emph{Spitzer} Space Telescope (Dunham et~al. 2008; Evans et~al. 2009) have aggravated the ``luminosity problem'' first articulated by Kenyon et~al. (1990). The trend is increasing numbers of protostars at lower luminosities down to an internal luminosity, which describes the total luminosity of the central protostar and circumstellar disc, of $L_{\rm{int}}\simeq 0.1\,\rm{L}_{\odot}$, below which the numbers start to decline. For young embedded protostars the accretion luminosity will dominate over that from the stellar photosphere. However, simple star formation models, such as the spherical accretion model of Shu (1977), predict that a low mass source on the stellar/brown dwarf boundary ($M\simeq0.08\,\rm{M}_{\odot}$) should have an internal luminosity $L_{\rm{int}}\simeq 1.6\,\rm{L}_{\odot}$ from accretion alone (Evans et~al. 2009). Non-steady accretion, starting in the earliest protostellar stages, is currently the best solution to this discrepancy (Kenyon \& Hartmann 1995; Young \& Evans 2005; Enoch et~al. 2007). The luminosity problem is most difficult to rectify in very low luminosity objects (VeLLOs; Young et~al. 2004; Dunham et~al. 2008) with extreme luminosities $L_{\rm{int}}\leq 0.1\,\rm{L}_{\odot}$. The nature of these objects is unclear, whether they are young Class O protostars which are just powering up, or are more evolved but in a low accretion state (Dunham et~al. 2008; Evans et~al. 2009).

The VeLLO IRAM~04191-IRS (Andr{\'e} et~al. 1999) is a good example of this situation. IRAM~04191-IRS has a well measured molecular outflow, from which the mass accretion onto the protostar can be calculated (Andr{\'e} et~al. 1999; Dunham et~al. 2008). The luminosity expected from this accretion, using steady accretion arguments, is found to be 25 times the measured value (Dunham et~al. 2006). However, it is not the case that all VeLLOs are so well defined, and indeed many of their physical properties are often inconsistent (Bourke et~al 2006). The existing sample of VeLLO sources, although expanded by \emph{Spitzer} (Dunham et~al. 2008), is in no way complete. VeLLOs are difficult to confirm as protostars in the infra-red due to their low luminosity and embedded nature, and measuring their molecular outflows may also be problematic as exemplified by the case of L1014 (Bourke et~al. 2005; Crapsi et~al. 2005). It is also the case that VeLLOs are often found in cores which are not only assumed to be starless, but which were also not believed to be approaching collapse (Bourke et ~al. 2006). Nevertheless, identifying a complete sample of these low luminosty embedded protostars is vital for understanding low mass star formation.

This paper presents high resolution, high sensitivity observations towards a sample of known or proposed protostellar objects all located within clouds targeted by a joint Arcminute Microkelvin Imager (AMI) and Sunyaev--Zel'dovich Array (SZA; Carlstrom et~al. ***) spinning dust dark cloud sample (Scaife et~al. in prep) which was selected from the \emph{Spitzer} small clouds and cores programme (Evans et~al. 2003) and is designed to measure the emission from the clouds on arcminute scales. The majority of the cores observed here have low bolometric luminosity, with a significant proportion being Very Low Luminosity Objects. Since these studies are on differing scales but target the same regions we now explicitly define our nomenclature. \emph{Cores} are defined as compact objects on scales of a few tenths of a parsec and are typically identified at sub-mm wavelengths. Cores can be protostellar or starless, a distinction made on the basis of their SEDs, the presence of an IR source, a molecular outflow or a compact cm-wave source (Andr{\'e} et~al. 2000). VeLLOs are a subset of cores, which until recently were often presumed to be starless but are in fact a low luminosity embedded population. Cores identified in the sub-mm are suffixed ``SMM'', whilst those identified in the infra-red are suffixed ``IRS''. \emph{Clouds} or \emph{dark clouds} are regions of high visual extinction on scales of a few parsecs. They contain large quantitities of molecular gas, which often harbours cores and active star formation. On larger scales of a few parsecs to a few tens of parsecs we refer to extended regions of presumably interconnected clouds as \emph{cloud complexes}.

The organization of this paper is as follows. In \S~\ref{sec:sample} we describe the sample of targets to be observed, and in \S~\ref{sec:obs} we describe the AMI Large Array (AMI-LA) telescope, the observations and the data reduction process. In \S~\ref{sec:results} we comment upon the results of the observations and compare them to predictions. Detailed notes on individual fields are given in \S~\ref{sec:notes}. The nature of the radio emission is discussed in \S~\ref{sec:disc} and correlations with the IR properties of the sample are derived in \S~\ref{sec:corr}. Finally we present our conclusions in \S~\ref{sec:conc}.

\section{Sample}
\label{sec:sample}

The thirteen fields observed by the AMI-LA are listed in Table~\ref{tab:lynds1}. These fields were selected from the full AMI-SZA spinning dust sample (Scaife et~al. in prep) on the basis of being known to contain cores, or alternatively to contain possible embedded objects from the \emph{Spitzer} catalog of Dunham et~al. (2008). Where necessary we shall identify candidates from this catalog by their catalog number, i.e. [DCE08]-nnn. These candidates were ranked by Dunham et~al. (2008) as belonging to one of six ``groups'', with Group 1 being those most likely to be true embedded objects, and Group 6 those least likely. Table~\ref{tab:lynds1} lists the fields in order of increasing Right Ascension with the associated cloud named in Column [1] . In Columns [2] and [3] it lists the pointing center observed for each field, and in Column [4] the association of the cloud to a cloud complex. In Columns [5--7] the existing sub-mm observations are listed, the references for which are given in Column [8]. Column [9] lists the number of cores of varying class thought to be in each field. Within the AMI-LA fields there are 21 known cores or potential embedded objects. Five fields contain known VeLLO objects: IRAM~04191 (Dunham et~al. 2006), L1521F (Bourke et~al. 2006; Terebey et~al. 2009; Shinnaga et~al. 2009), L673-7 (Dunham et~al. 2008), L1148 (proposed; Kauffmann et~al. 2005), and L1014 (Young et~al. 2004; Bourke et~al. 2005; Shirley et~al. 2007). To this list we add two further potential VeLLO objects: L723-IRS ([DCE08]-026) and L1165-IRS ([DCE08]-039), which were included in the catalog of Dunham et~al. (2008) but ranked as Group 6.

All of the fields have been at least partially observed in the sub-mm, although three of the VeLLO objects (L723-IRS and L1165-IRS) are not covered by existing observations. Several fields also have data at lower radio frequencies: IRAM~04191 (Andr{\'e}, Motte \& Bacmann 1999), L1521F (Harvey et~al. 2002), B35A (Rodr{\'i}guez et~al. 1989), L723 (Anglada et~al. 1991; Carrasco-Gonz{\'a}lez et~al. 2008), L1014 (Shirley et~al. 2007) and L1221 (Rodr{\'i}guez \& Reipurth 1998; Young et~al. 2009). The correspondance of the AMI-LA measurements to these observations are discussed in more detail in \S~\ref{sec:fcorr}.

\section{Observations}
\label{sec:obs}

AMI comprises two synthesis arrays, one of ten 3.7\,m
antennas (Small Array; SA) and one of eight 13\,m antennas (Large Array; LA),
both sited at Lord's Bridge, Cambridge (AMI Consortium: Zwart
et~al. 2008). The telescope observes in 
the band 13.5--17.9\,GHz with cryostatically cooled NRAO indium--phosphide
front-end amplifiers. The overall system temperature is approximately
25\,K. Amplification, equalization, path compensation and
automatic gain control are then applied to the IF signal. The backend has an analogue lag correlator with 16 independent
correlations formed at twice Nyquist rate for each baseline using path delays spaced by
25\,mm. From these, eight complex visibilities are formed in
0.75\,GHz bandwidth channels. In practice, the two
lowest frequency channels (1 \& 2) are not generally used due to a lower response in this frequency range and interference from geostationary
satellites. The data in this paper were taken with the AMI Large Array.

Observations of the thirteen fields listed in Table~\ref{tab:lynds1} were made with the AMI-LA between October 2009 and July 2010. The clouds were observed as single pointings, with the exception of B35A and L1148. B35A was observed as a 7 field mosaic in order to cover the same sky area as earlier VLA observations, see \S~\ref{sec:notes}, and L1148 was observed in two separate pointings as the two embedded cores proposed by Dunham et~al. (2008) for this cloud are separated by more than the AMI-LA primary beam FWHM.

AMI-LA data reduction is performed using the local software tool \textsc{reduce}. This applies
both automatic and manual flags for interference, 
shadowing and hardware errors, Fourier transforms the correlator data to synthesize frequency
channels and performs phase and amplitude
calibrations before output to disc in \emph{uv} FITS format suitable for imaging in
\textsc{aips}. Flux (primary) calibration is performed using short observations of 3C286 and 3C48. We assume I+Q flux densities for this source in the
AMI LA channels consistent with the updated VLA calibration scale (Rick Perley, private comm.), see Table~\ref{tab:cals}. Since the AMI-LA measures
I+Q, these flux densities 
include corrections for the polarization of the calibrator sources. A correction is
also made for the changing intervening air mass over the observation. From
other measurements, we find the flux calibration is accurate to better than
5 per cent (AMI Consortium: Scaife et~al. 2008; AMI Consortium:
Hurley--Walker et~al. 2009). Additional phase (secondary) calibration is done using interleaved observations of
calibrators 
selected from the Jodrell Bank VLA Survey (JVAS; Patnaik et~al. 1992). After calibration, the phase is generally stable to
$5^{\circ}$ for channels 4--7, and
$10^{\circ}$ for channels 3 and 8. The FWHM of the primary beam of the AMI LA is $\approx 6$\,arcmin at 16\,GHz. 

Reduced data were imaged using the AIPS data package. {\sc{clean}}
deconvolution was performed using the task 
{\sc{imagr}} which applies a differential primary beam correction to
the individual frequency channels to produce the combined frequency
image.  The broad
spectral coverage of AMI allows a representation of the spectrum
between 14.3 and 17.9\,GHz to be made independently of other
telescopes when sufficient signal to noise is present and in what
follows we use the convention: $S_{\nu}\propto \nu^{\alpha}$, where $S_{\nu}$ is
flux density (rather than flux, $F_{\nu}=\nu S_{\nu}$), $\nu$ is frequency and $\alpha$ is the spectral index. All errors quoted are 1\,$\sigma$. 

The observations towards the thirteen fields listed in Table~\ref{tab:lynds1} and described in \S~\ref{sec:sample} are summarized in Table~\ref{tab:lynds2}. The details of each individual observation including the date and both the primary and secondary calibration sources are listed, along with the resulting rms noise level in the map and the dimensions of the naturally weighted synthesized beam. The rms noise level varies between fields due to the different levels of data flagging required following periods of poor weather conditions or interference from non-astronomical sources, such as geostationary satellites. Where possible for observations that were heavily flagged, a second observation was made and these are also indicated in Table~\ref{tab:lynds2}.

\begin{table*}
\caption{AMI-LA Fields. Column [1] Name of field, [2] Right Ascension, [3] Declination, [4] Association, [5] flag on existing data from SCUBA, [6] flag on existing data from MAMBO, [7] flag on existing data from SHARC-II, [8] references for [5--7], [9] number of cores (confirmed and proposed) within the area covered by the AMI-LA observation.  \label{tab:lynds1}}
\begin{tabular}{lcccccccc}
\hline \hline
Name & RA  & Dec  & Association &  SCUBA & MAMBO & SHARC-II & refs. & $N_{\rm{core}}$\\
     & (J2000) & (J2000) & & Y/N & Y/N & Y/N & & \\
\hline
IRAM04191 & 04 21 57 & 15 29 46 & Taurus & Y & N & Y & 1,6 & 2\\
L1521F    & 04 28 39 & 26 51 36 & Taurus &   Y & Y & Y & 1,5,6 & 1 \\
B35A      & 05 44 29 & 09 08 57 & - &  Y & N & Y & 1,5,6 & 2\\
L673      & 19 20 26 & 11 22 19 & Aquila Rift, Cloud B & Y & N & N & 2 & 3\\
CB188	  & 19 20 15 & 11 36 08 & Aquila Rift, Cloud B & N & Y & Y & 4,5,6 & 1\\
L673-7    & 19 21 35 & 11 21 23 & Aquila Rift, Cloud B  & N & Y & Y & 5,6 & 1\\
L723      & 19 17 44 & 19 15 24 & - & Y & N & Y & 1,6 & 2\\
L1152     & 20 35 46 & 67 53 02 & - & Y & N & Y & 1,6 & 1\\
L1148	  & 20 40 57 & 67 23 05 & Cepheus Flare & N & Y & Y & 3,5,6 & 2\\
BERN 48   & 20 59 15 & 78 22 60 & Cepheus Flare & Y & Y & Y & 5,6 & 1\\
L1014     & 21 24 08 & 49 59 09 & - & Y & Y & Y & 2,5,6 & 1\\
L1165     & 22 06 50 & 59 02 46 & Cloud 157, Cyg OB7 & N & N & Y & 2,6 & 2\\
L1221     & 22 28 07 & 69 00 39 & - & Y & N & Y & 1,6 & 2 \\
\hline
\end{tabular}
\begin{minipage}{\textwidth}{ 
[1] Young et~al. 2006;
[2] Visser et~al. 2002;
[3] Kirk, Ward-Thompson \& Andr{\'e} 2007;
[4] Launhardt et~al. 1997;
[5] Young et~al. 2006b;
[6] Wu et~al. 2007.
}
\end{minipage}
\end{table*}
\begin{table}
\caption{AMI-LA frequency channels and primary calibrator flux densities measured in Jy.\label{tab:cals}}
\begin{tabular}{lcccccc}
\hline \hline
Channel No. & 3 & 4 & 5 & 6 & 7 & 8\\
Freq. [GHz] & 14.27 & 14.99 & 15.71 & 16.43 & 17.15 & 17.87 \\
\hline
3C48  & 1.85 & 1.75 & 1.66 & 1.58 & 1.50 & 1.43 \\
3C286 & 3.60 & 3.54 & 3.42 & 3.31 & 3.21 & 3.11 \\
3C147 & 2.75 & 2.62 & 2.50 & 2.40 & 2.30 & 2.20 \\
\hline
\end{tabular}
\end{table}
\begin{table*}
\caption{AMI-LA Observations. Column [1] Name of field, [2]
Date of initial observation, [3] Primary Calibrator, [4] Secondary Calibrator$^a$, [5] Date of additional observation, [6] Primary Calibrator, [7] Secondary Calibrator$^a$, [8] AMI-LA combined data synthesized beam FWHM
major axis, [9] AMI-LA combined data synthesized beam FWHM minor axis, and [10]
rms noise fluctuations on the combined channel map. \label{tab:lynds2}}
\begin{tabular}{lccccccccc}
\hline \hline
Name & Date  & $1^{\circ}$ & $2^{\circ}$ & Date & $1^{\circ}$ & $2^{\circ}$ & $\Delta \theta_{\rm{maj}}$ & $\Delta
\theta_{\rm{min}}$ & $\sigma_{\rm{rms}}$\\
     & (dd-mm-yy) & & & (dd-mm-yy) & & & (arcsec) & (arcsec) &
($\frac{\mu{\rm{Jy}}}{\rm{bm}}$)\\
\hline
IRAM04191 & 09-10-09 & 3C48 & J0424+1442 & 07-11-09 & 3C286 & J0424+1442 & 49.9 & 27.4 & 17 \\
L1521F    & 17-10-09 & 3C147& J0426+2350 & - & - & - & 42.0 & 26.3 & 16\\
B35A$^b$  & 08-11-09 & 3C48 & J0551+0829 & - & - & - & 53.6 & 28.0 & 67 \\
L673      & 13-12-09 & 3C48 & J1922+1530 & - & - & - & 36.5 & 31.2 & 31 \\
CB188	  & 27-12-09 & 3C48 & J1922+1530 & - & - & - & 46.0 & 21.9 & 24\\
L673-7    & 18-10-09 & 3C286 & J1922+1530 & - & - & - & 32.0 & 26.3 & 16 \\
L723      & 14-12-09 & 3C286 & J1905+1943 & 08-10-09 & 3C286 & J1925+2106 & 42.9 & 24.0 & 29 \\
L1152     & 26-12-09 & 3C286 & J2035+5821 & 05-01-10 & 3C286 & J2035+5821 & 38.8 & 22.6 & 49\\
L1148$^c$ & 12-10-09 & 3C286 & J2052+6858 & 20-10-09 & 3C286 & J2052+6858 & 36.6 & 26.5 & 22 \\
BERN 48   & 15-11-09 & 3C48 & J2051+7441 & - & - & - & 30.0 & 27.4 & 23 \\
L1014     & 14-10-09 & 3C48 & J2123+4614 & - & - & - &31.2 & 25.4 & 21\\
L1165     & 30-06-10 & 3C147 & J2223+6249 & 02-07-10 & 3C48 & J2223+6249 & 32.3 & 29.0 & 21 \\
L1221     & 11-10-09 & 3C286 & J2230+6946 & - & - & - & 27.4 & 25.1 & 19 \\
\hline
\end{tabular}
\begin{minipage}{\textwidth}{
$^a$ Secondary calibrators are selected from the JVAS catalog (Patnaik et~al. 1992).

$^b$ The 7 pointings of the B35A cloud mosaic were observed in an interleaved fashion during single run.

$^c$ The 2 pointings towards the L1148 cloud were observed on separate dates, the later observation corresponds to the southern pointing.
}
\end{minipage}
\end{table*}

\section{Results}
\label{sec:results}

Within the 13 fields observed by the AMI-LA we detect 40 sources, and we identify 15 of these as being associated with the 21 possible cores in these areas. The combined channel maps from the AMI-LA are shown in Figs.~\ref{fig:figs1}--\ref{fig:figs3}. The rms noise on each map, $\sigma_{\rm{rms}}$, is found using the {\sc aips} task {\sc imean} which fits a Gaussian profile to the histogrammed pixel values for each field. These values may be found in Table~\ref{tab:lynds2} and in the figure captions. We identify all objects with peak flux densities greater than $5\,\sigma_{\rm{rms}}$, where that peak lies within the FWHM of the AMI-LA primary beam, as being true sources and these are listed in Table~\ref{tab:fluxes}. The two detection conditions are relaxed for some individual cases and where this occurs it is indicated in Table~\ref{tab:fluxes} and commented upon in Section~\ref{sec:notes}. 

Table~\ref{tab:fluxes} lists sources in order of increasing Right Ascension for each field, with peak positions given in Columns [3] and [4]. Given the low signal-to-noise in the individual channel data for these targets, the reduced phase stability of AMI-LA Channel 3 was found to produce an rms noise too high relative to the other channels to contribute usefully to the combined dataset, consequently it has not been used in this work. In addition, since channel 8 of the AMI-LA is prone to satellite interference it is sometimes not used in making the final combined bandwidth images. As a rule, this channel is only included when the resulting noise level in the combined channel map, $\sigma_{\rm{rms}}^{4-8}$, is lower than that of a map using only channels 4 to 7, $\sigma_{\rm{rms}}^{4-7}$. As this varies from field to field, changing the nominal observation frequency slightly, Column [5] lists this frequency for each dataset. For ease of notation AMI-LA measurements will henceforth be denoted as being made at 16\,GHz (1.8\,cm) when described in the text; for the more precise measurement frequency of individual fields we refer the reader to Table~\ref{tab:fluxes}. Column [6] lists the primary beam corrected peak flux density for each detected source with no correction for the local background, and Column [7] the primary beam corrected integrated flux density. Integrated flux densities are found using the {\sc fitflux} program (Green 2007). This method calculates flux densities by removing a tilted
plane fitted to the local background and integrating the remaining
flux density. This is done by drawing a polygon around the source and fitting
a tilted plane to the pixels around the edge of the polygon. Where an
edge of the polygon crosses a region confused by another source, the
background is subjective and this edge is omitted from the fitting. Since the extracted flux density is dependent to some degree on the
background emission, we repeat this process using five irregular
polygons, each varying slightly in shape. The final flux density is
the average of that extracted from these five polygons. The error on the integrated flux densities is calculated as $\sigma_{\rm{S}}=\sqrt{(0.05 S_{\rm{int}})^2 + \sigma_{\rm{fit}}^2 +\sigma_{\rm{rms}}^2}$, where $\sigma_{\rm{fit}}$ is the standard deviation of the flux densities found from the five polygonal apertures, $\sigma_{\rm{rms}}$ is the rms noise determined using {\sc imean} and $0.05 S_{\rm{int}}$ is a conservative 5~per~cent absolute calibration error. Where the 5~per~cent calibration uncertainty is not dominant, errors determined for this sample are in general heavily dependent on the rms noise, $\sigma_{\rm{rms}}$, rather than the fitting error, $\sigma_{\rm{fit}}$, which is found to be small. Column [8]  of Table~\ref{tab:fluxes} then identifies which of the detected sources are associated with known cores.

\subsection{A Note on Expected Sources}

At 16\,GHz we expect a certain number of extragalactic radio sources to be seen within each of our fields. To quantify this number we use the 15\,GHz source counts model from de~Zotti et~al. (2005) scaled to the extended 9C survey source counts (Waldram et~al. 2010). The average rms noise from our datasets is $\simeq25\,\mu$Jy\,beam$^{-1}$ and from this model we predict that we should see 0.07 sources\,arcmin$^{-2}$, $\approx 2$ radio sources within a 6\,arcmin FWHM primary beam above a $5\,\sigma$ flux density of 125\,$\mu$Jy, or $31\pm6$ sources in total. On average in our sample we see 0.09 sources arcmin$^{-2}$. Since we identify 15 of our 40 sources with protostellar cores, not accounted for by the extragalactic radio source counts model, this is consistent with the discrepancy in the measured and predicted source counts. Since the short frequency coverage of the AMI-LA band makes spectral index estimates highly uncertain for low signal to noise detections, and given this statistic, we tentatively identify any sources found in the target fields that do not have a known protostellar association as extragalactic.

\begin{table*}
\caption{Sources detected within the AMI-LA fields. \label{tab:fluxes}}
\begin{tabular}{llcccccc}
\hline\hline
Field & Source & RA & Dec & $\bar{\nu}$ &$S_{\rm{peak}}$ & $S_{\rm{int}}$ & assoc. \\
&&(J2000)&(J2000)& (GHz) &($\mu$Jy\,beam$^{-1}$)& ($\mu$Jy) &\\
\hline
IRAM04191 &(1)& 04 21 56.7 & 15 29 36.0 & 16.43 & 164.3  & 172.9$\pm$20.1 & IRAM~04191-IRS\\
	  &(2)& 04 22 05.0 & 15 28 01.0 & 16.43 & 185.5  & 316.7$\pm$27.1 &\\
L1521F	  &(1)& 04 28 36.4 & 26 53 06.0 & 16.07 & 120.0  & 90.5$\pm$17.9  &\\
	  &(2)& 04 28 38.6 & 26 51 11.0 & 16.07 & 132.9  & 139.2$\pm$18.2 & L1521F-IRS\\
	  &(3)& 04 28 40.9 & 26 51 51.0 & 16.07 & 138.8  & 234.2$\pm$22.0 & L1521F-IRS\\
	  &(4)& 04 28 41.2 & 26 53 56.0 & 16.07 & 5822.6 & 6136.1$\pm$308.4  &\\
	  &(5)& 04 28 45.4 & 26 53 56.0 & 16.07 & 247.4  & 289.8$\pm$19.9 &\\
	  &(6)& 04 28 47.2 & 26 52 36.0 & 16.07 & 117.5  & 139.1$\pm$18.7 &\\
B35A      &(1$^{\ast}$)& 05 44 06.0 & 09 08 42.0 & 16.07 & 1110.8 & 936.1$\pm$82.1 &\\
   	  &(2$^{\ast}$)& 05 44 08.3 & 09 08 02.0 & 16.07 & 1440.3 & 1433.5$\pm$100.1 &\\
   	  &(3)& 05 44 19.5 & 09 11 17.0 & 16.07 & 352.3 & 363.7$\pm$70.2 &\\ 
   	  &(4)& 05 44 22.5 & 09 09 47.0 & 16.07 & 585.6 & 558.2$\pm$73.1 &\\
   	  &(5)& 05 44 29.6 & 09 08 52.0 & 16.07 & 591.7 & 943.0$\pm$84.5 & B35A~SMM-1\\
   	  &(6)& 05 44 31.0 & 09 07 42.0 & 16.07 & 355.7 & 683.2$\pm$78.4 & \\
   	  &(7)& 05 44 43.8 & 09 03 52.0 & 16.07 & 811.3 & 1139.2$\pm$92.3  &\\
   	  &(8)& 05 44 48.5 & 09 03 27.0 & 16.07 & 7450.2 & 8007.4$\pm$407.3 &\\
   	  &(9)& 05 44 53.2 & 09 06 22.0 & 16.07 & 676.7 & 642.3$\pm$75.4 &\\
   	  &(10)&05 44 53.9 & 09 10 17.0 & 16.07 & 389.2 & 430.9$\pm$76.4 &\\
   	  &(11)&05 45 02.0 & 09 05 01.9 & 16.07 & 394.1 & 343.7$\pm$70.1 &\\
L673	  &(1)& 19 20 25.0 & 11 22 15.0 & 16.07 & 354.6 & 455.2$\pm$39.3 & L673~SMM-1\\
	  &(2)& 19 20 26.0 & 11 20 15.0 & 16.07 & 218.4 & 289.7$\pm$35.9 & L673~SMM-2\\
	  &(3)& 19 20 35.2 & 11 21 00.0 & 16.07 & 300.6 & 380.3$\pm$39.0 &\\
CB188	  &(1$^{\ddagger}$)& 19 20 14.7 & 11 35 48.0 & 16.43 & 179.5 & 179.5$^{\ddagger}$ & CB188~SMM-1\\
L673-7    &(1)& 19 21 28.9 & 11 22 48.0 & 16.07 & 163.6 & 548.1$\pm$37.5 &\\
	  &(2)& 19 21 33.0 & 11 22 48.0 & 16.07 & 184.6 & 428.4$\pm$34.3 &\\
L723      &(1)& 19 17 44.4 & 19 15 24.0 & 16.43 & 122.2 & 220.0$\pm$32.0 & L723-IRS\\
          &(2)& 19 17 46.5 & 19 14 29.0 & 16.43 & 127.3 & 154.0$\pm$30.8 & \\
          &(3$^{\ast}$)& 19 17 53.5 & 19 12 29.0 & 16.43 & 610.0 & 689.5$\pm$45.5 & L723~SMM-1\\
L1152	  &(1)& 20 35 29.2 & 67 52 07.0 & 16.43 & 2470.2 & 2471.5$\pm$133.1 &  \\
L1148     &(1)& 20 40 41.4 & 67 25 20.0 & 16.07 & 116.7 & 205.5$\pm$26.3 &\\
	  &(2$^{\dagger}$)& 20 40 57.0 & 67 22 55.0 & 16.07 & 83.7 & 71.9$^{\dagger}$ & L1148-IRS\\
BERN~48	  &(1)& 20 59 15.0 & 78 23 04.9 & 16.43 & 597.4  & 615.2$\pm$38.7 & Bern~48\\
L1014     &(1)& 21 23 55.0 & 49 58 59.0 & 16.07 & 1615.6 & 1680.2$\pm$87.1  &\\
	  &(2)& 21 23 56.1 & 49 57 54.0 & 16.07 & 203.3  & 249.7$\pm$28.7 &\\
	  &(3)& 21 24 08.5 & 49 59 04.0 & 16.07 & 166.3  & 298.5$\pm$26.5 & L1014-IRS\\
	  &(4)& 21 24 17.3 & 50 01 04.0 & 16.07 & 242.7  & 229.8$\pm$28.8 &\\
	  &(5)& 21 24 20.4 & 49 57 24.0 & 16.07 & 682.3  & 634.0$\pm$38.7 &\\
L1165     &(1)& 22 06 17.5 & 59 04 29.9 & 16.43 & 250.5  & 397.1$\pm$33.0 & \\
	  &(2)& 22 06 30.5 & 59 02 40.0 & 16.43 & 109.9  & 81.6$\pm$22.8  & \\
	  &(3)& 22 06 35.0 & 59 02 35.0 & 16.43 & 186.4  & 346.4$\pm$30.6 & L1165-IRS\\
	  &(4)& 22 06 49.9 & 59 03 00.0 & 16.43 & 240.7  & 305.8$\pm$28.2 & L1165~SMM-1\\
L1221     &(1)& 22 27 43.7 & 69 00 48.9 & 16.43 & 1535.5 & 1704.3$\pm$87.7  &\\
	  &(2)& 22 27 51.2 & 69 00 24.0 & 16.43 & 187.9  & 182.2$\pm$24.2 &\\
    	  &(3)& 22 27 52.1 & 69 01 39.0 & 16.43 & 1867.5 & 2081.7$\pm$106.4  &\\	
	  &(4)& 22 28 03.3 & 69 01 14.0 & 16.43 & 282.8  & 358.8$\pm$27.6 & L1221-IRS1\\
	  &(5)& 22 28 07.0 & 69 00 39.0 & 16.43 & 367.4  & 358.9$\pm$28.1 & L1221-IRS3\\
	  &(6)& 22 28 35.8 & 69 00 08.9 & 16.43 & 350.9  & 353.9$\pm$27.8 &\\
\hline
\end{tabular}
\begin{minipage}{\textwidth}{$^{\ast}$These sources are outside the AMI-LA primary beam FWHM.\\
$^{\dagger}$These sources are detected at $3\,\sigma$.\\
$^{\ddagger}$Lower limit on flux density, see text for details.
}
\end{minipage}
\end{table*}

\section{Notes on Individual Fields}
\label{sec:notes}

\subsection{IRAM04191} 
\begin{figure*}
\centerline{\includegraphics[width=0.4\textwidth]{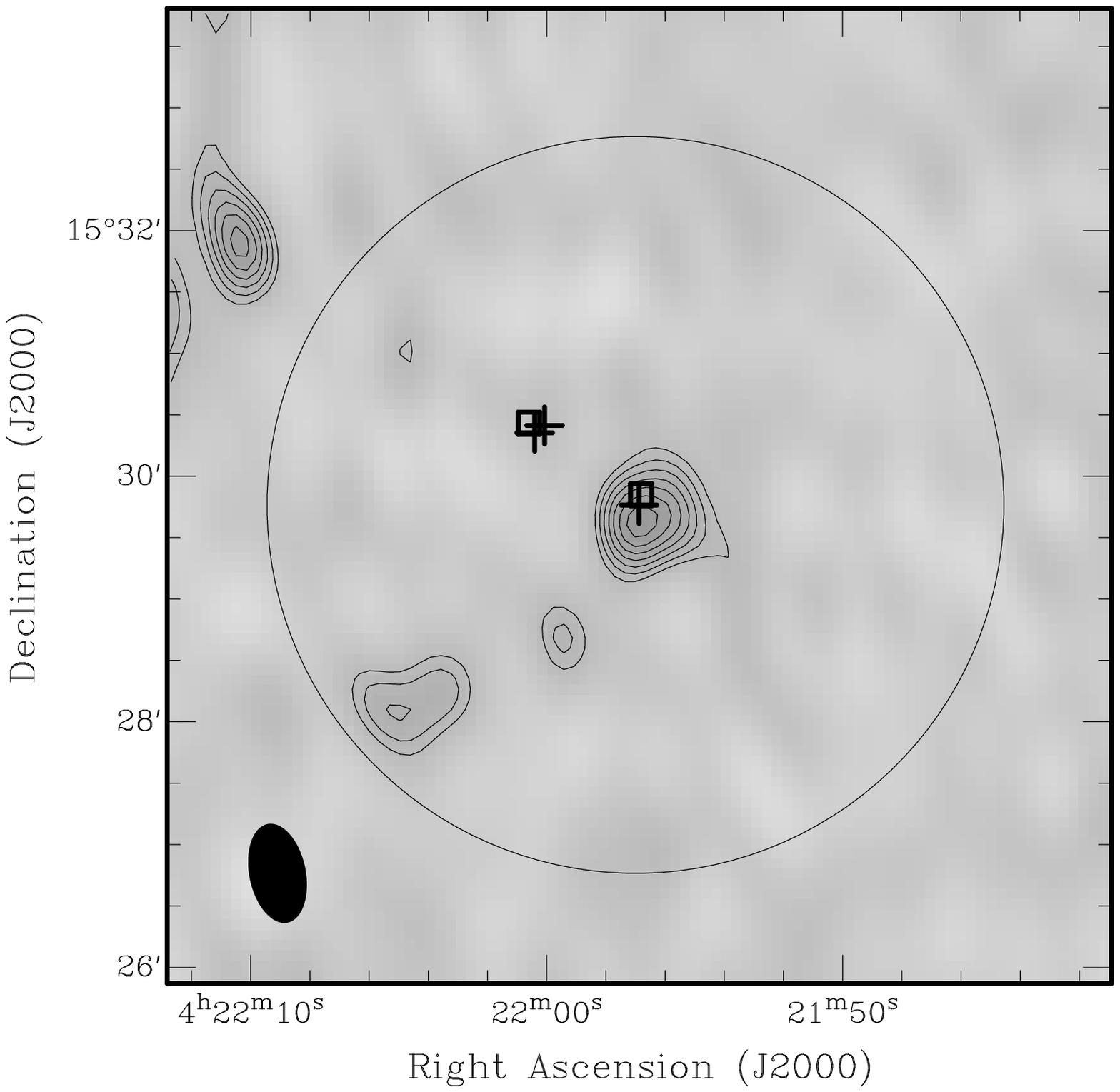}\qquad\includegraphics[width=0.4\textwidth]{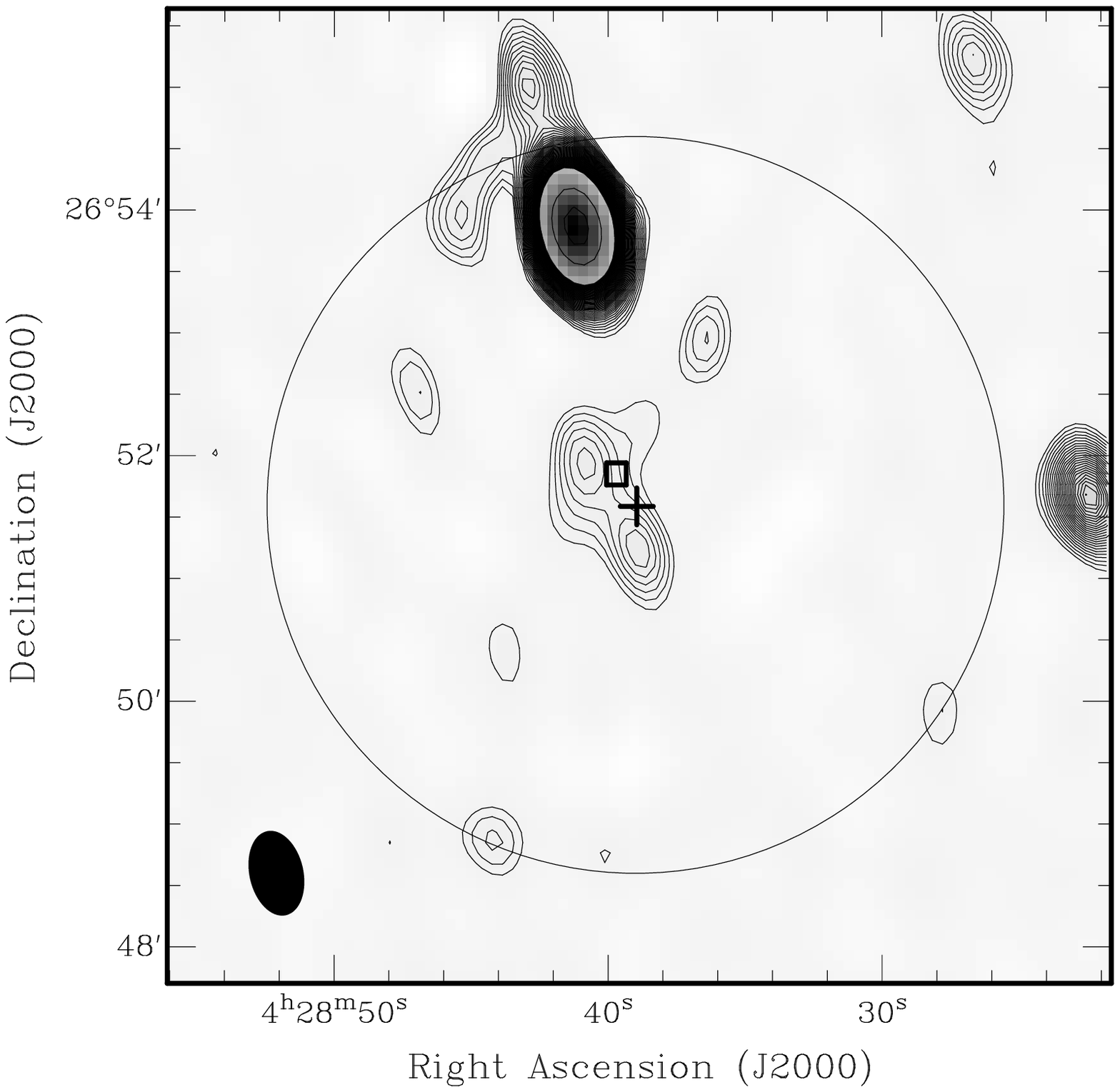}}
\centerline{(a)\hspace{0.4\textwidth}(b)}
\centerline{\includegraphics[width=0.4\textwidth]{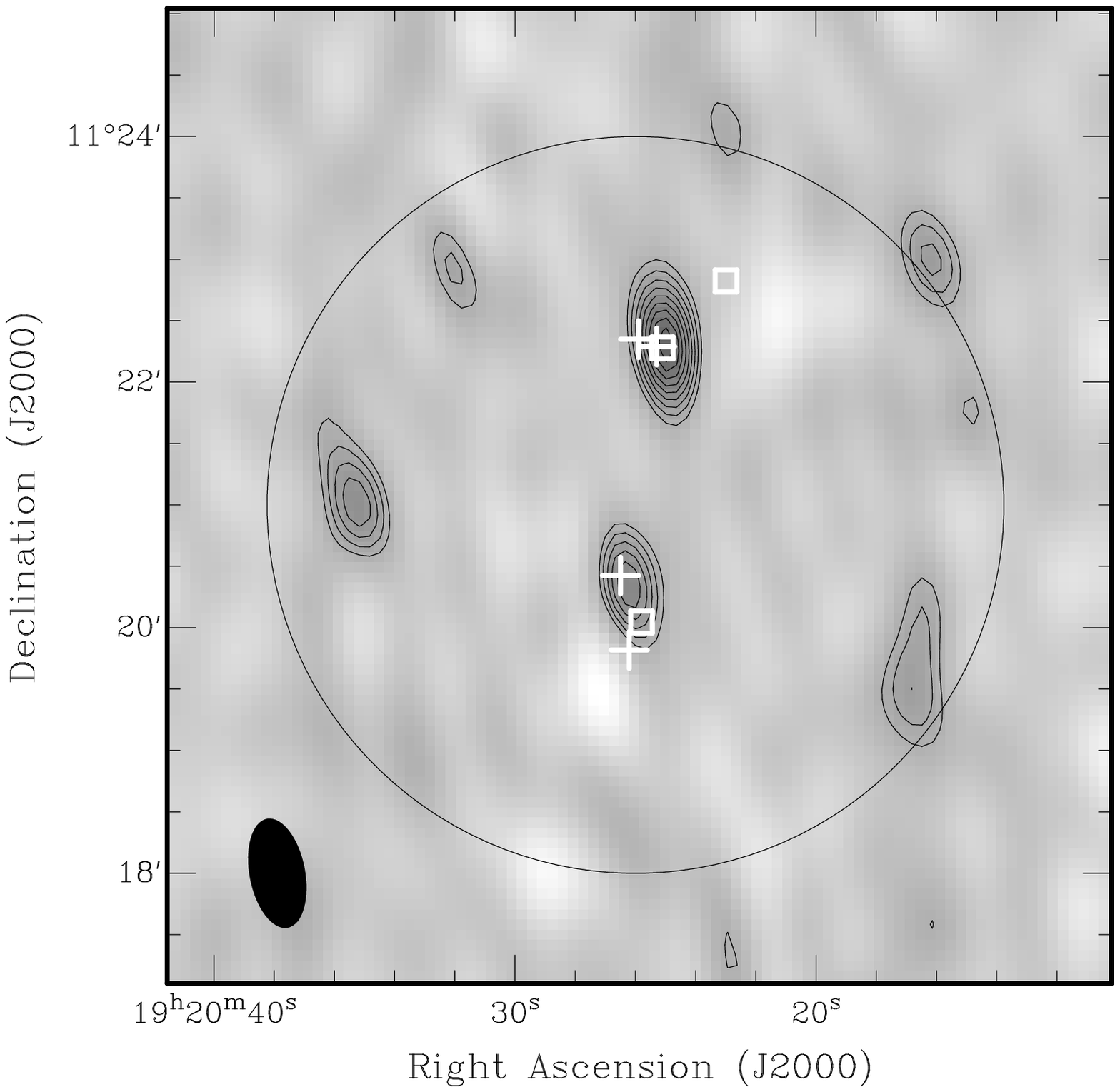}\qquad\includegraphics[width=0.4\textwidth]{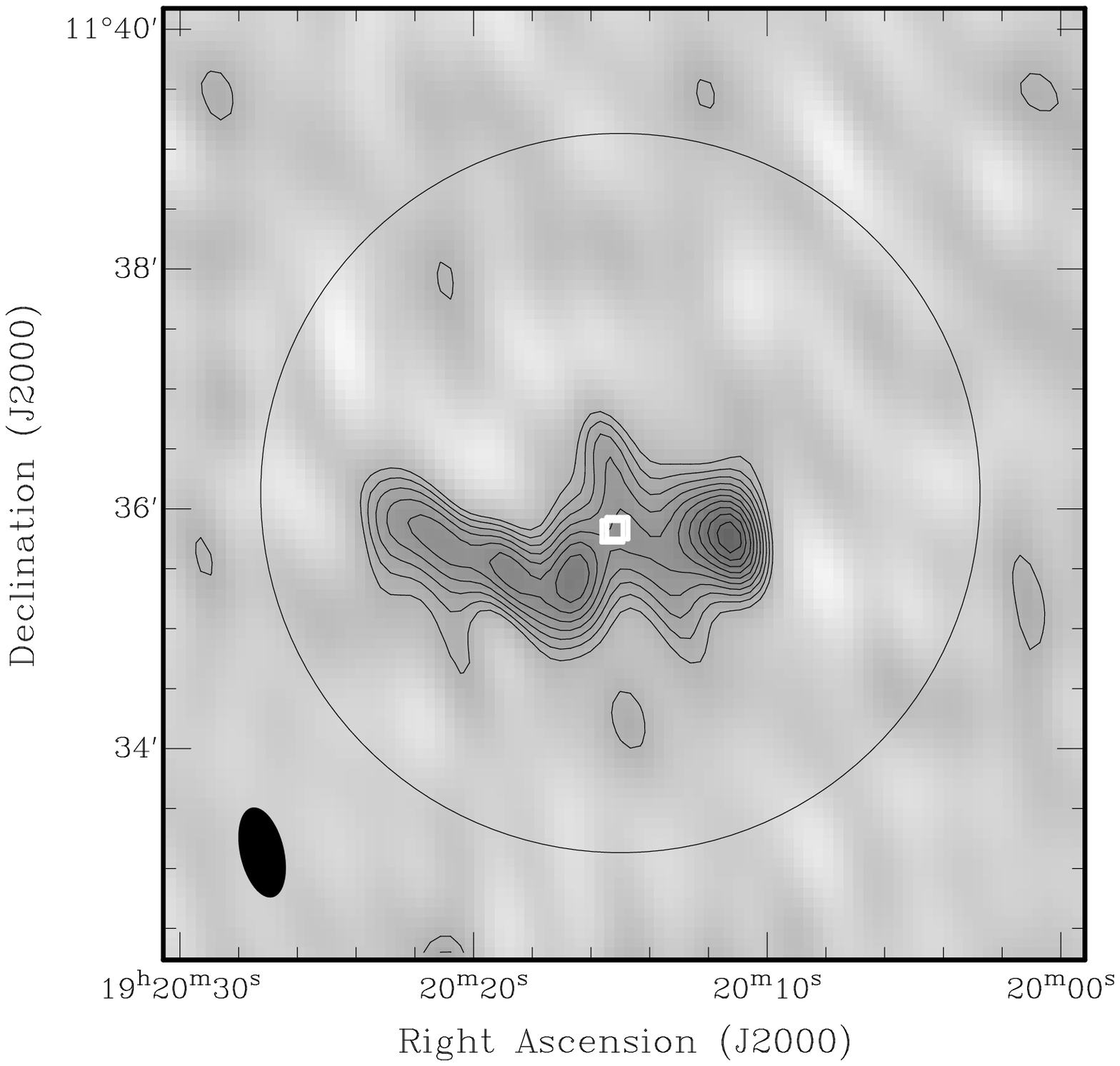}}
\centerline{(c)\hspace{0.4\textwidth}(d)}
\caption{(a) IRAM~04191; (b) L1521F; (c) L673; and (d) CB188. AMI-LA data are shown as greyscale and contours linearly from 3\,$\sigma$, where $\sigma=17, 16, 26$ and $24\,\mu$Jy\,beam$^{-1}$ respectively.  \emph{Spitzer} positions and candidate embedded objects from Dunham et~al. (2008) are indicated by crosses, and the positions of sub-mm cores are shown as unfilled squares, see text for details. The AMI-LA primary beam FWHM is shown as a circle and the AMI-LA synthesized beam is shown as a filled ellipse in the lower left corner of each map. \label{fig:figs1}}
\end{figure*}

IRAM~04191+1522 (IRAM~04191-IRS), see Fig.~\ref{fig:figs1}\,(a), was identified as a Class O source by Andr{\'e}, Motte \& Bacmann (1999; AMB99). It has a large extended molecular outflow, seen in CO(2-1) (AMB99) and a flattened core mapped in N$_2$H$^+$ (Dunham et~al. 2006). AMB99 reported radio flux densities for this object at 5, 8 and 15\,GHz from the VLA, which they stated were consistent with a spectral index of $\alpha=0.6$, similar to that seen in the radio from VLA1623. We find a flux density of $S_{16} = 0.173\pm0.020\,\mu$Jy for IRAM~04191-IRS, consistent with the earlier VLA measurement of $S_{15}^{\rm{VLA}} = 0.16\pm0.05\,\mu$Jy. Combining the VLA and AMI-LA channel data gives a spectral index of $\alpha_{4.85}^{16}=0.45\pm0.20$. Across the AMI-LA band alone we measure a spectral index of $\alpha_{14.9}^{17.8}=1.44\pm1.09$. The large error from the relatively small frequency coverage means that this is broadly consistent with that found at lower frequencies, but it may also indicate a steepening of the spectrum. The morphology of the object in the AMI-LA maps has a deconvolved major axis of $29''$ with a position angle of $112^{\circ}$ and is unresolved in the minor direction. This morphology is similar to that seen at 350\,$\mu$m (Dunham et~al. 2006). 
Also within the AMI-LA field, to the north-east of IRAM~04191-IRS, is the similarly named IRAS~04191, a Class I source. We detect no radio emission from this object. There is only one further source detected above 5\,$\sigma$ within the AMI-LA primary beam FWHM towards this field, and in the absence of further information we identify this as an extragalactic radio source. 

\subsection{L1521F} L1521F-IRS, see Fig.~\ref{fig:figs1}\,(b), is thought to be a core in the very earliest stages of gravitational collapse (Terebey et~al. 2009). Unlike the VeLLO sources IRAM04191-IRS or L1014-IRS it does not possess a detected molecular outflow. Previous deep cm-wave observations of the L1521F cloud (Harvey et~al. 2002), which covered L1521F-IRS, at 3.6\,cm wavelength did not find a coincident radio source and instead used a nearby object, which we identify as NVSS~042841+265355, to place an upper limit on the bolometric luminosity of the core. In Fig.~\ref{fig:figs1}\,(b) the source NVSS~042841+265355 can be seen to the north of the phase centre, just within the primary beam FWHM. This source has a primary beam corrected flux density of $S_{16}=6.1\pm0.3$\,mJy and in combination with $S_{1.4}=37.3\pm1.2$\,mJy (NVSS) and $S_{8.35}=10.8$\,mJy (Harvey et~al. 2002) we find this consistent with the presence of an extra-galactic non-thermal point source (as also concluded by Harvey et~al). However, in addition to NVSS~042841+265355 we also detect a two component radio structure more coincident with L1521F-IRS. Shinnaga et~al. (2009) detected extended warm gas (30-70\,K) towards L1521F-IRS and have suggested that it is highly likely this gas is heated through shocks produced by an outflowing/rotating component impacting on the surrounding cold dynamically collapsing medium. Such shocks would also be likely to produce radio emission from shock ionization and it may be that the double radio structure we see here is a consequence of L1521F-IRS's bipolar outflow, although we note that the radio emission is not aligned east--west following the morphology of the outflow as mapped by Bourke et~al. (2006). This may be a consequence of the non-uniformity of the surrounding interstellar medium, or as suggested by Shinnaga et~al. (2009) be due to the outflows having rotated from the position of the shocks responsible for the radio emission. Of the three further sources in this field, none have known radio or protostellar counterparts and in the absence of further information we identify them as extragalactic sources.

\subsection{B35A} 
\begin{figure*}
\centerline{\includegraphics[width=0.5\textwidth]{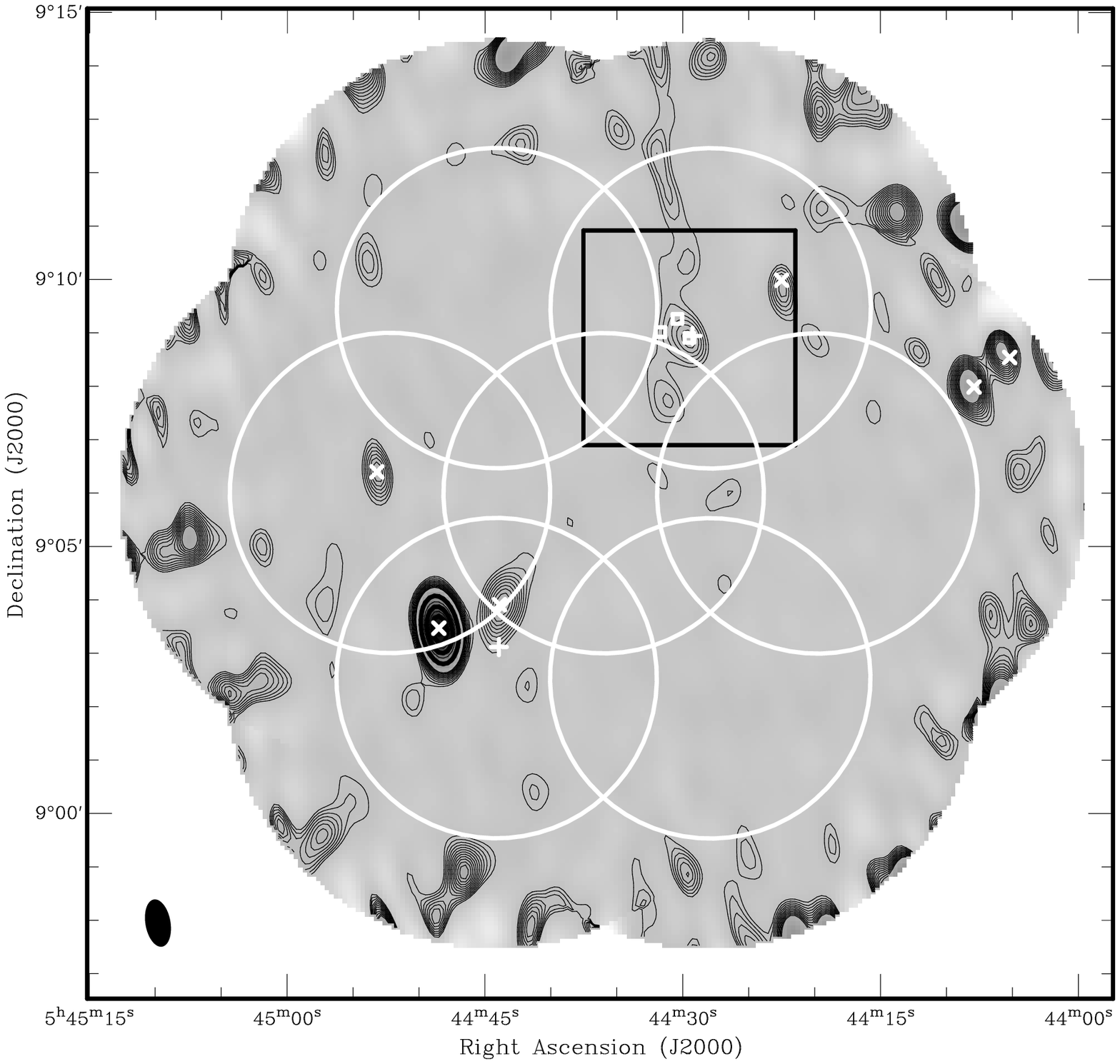}}
\caption{Primary beam corrected mosaic of B35A. AMI-LA data are shown as greyscale and contours linearly from 3\,$\sigma$, where $\sigma=67\,\mu$Jy\,beam$^{-1}$, to 1\,mJy\,beam$^{-1}$ and then incrementally in units of 1\,mJy\,beam$^{-1}$.  \emph{Spitzer} candidate embedded objects from Dunham et~al. (2008) are indicated by vertical crosses ($+$), and the positions of sub-mm cores are shown as unfilled squares, see text for details. The location of B35A~SMM-1 is highlighted by a square frame to the north-west of the centre. Radio sources identified by RR98 are marked by rotated crosses ($\times$). The AMI-LA primary beam FWHMs are shown as circles and the AMI-LA synthesized beam is shown as a filled ellipse in the lower left corner.   \label{fig:b35a}}
\end{figure*}

B35A~SMM-1, see Fig.~\ref{fig:b35a}, is an evolved protostar with observed molecular outflows (Myers et~al. 1988). A 6\,cm radio survey with the VLA (Rodr{\'i}guez et~al. 1998) placed an upper limit of 0.2\,mJy flux density on any radio emission from the protostar and we here observe a 7-field mosaic of the same sky area. At 16\,GHz we observe $0.94\pm0.08$\,mJy of flux density coincident with the B35A~SMM-1 core, as identified by the Submillimeter High Angular Resolution Camera II (SHARC-II; Wu et~al. 2007). Although the radio source extends slightly in their direction we do not see separate peaks from the adjacent cores B35A~SMM-2 and B35A~SMM-3. We detect all the radio sources identified at 6\,cm with the VLA and confirm that they are universally steep-spectrum objects and therefore most likely to be non-thermal extragalactic sources. The second of the candidate protostars proposed by Dunham et~al. (2008), which we denote B35A-IRS, is located in the most south-easterly AMI-LA pointing adjacent to a bright extragalactic radio source. Any emission from this source cannot be separately resolved by the AMI-LA synthesized beam. Of the three further sources detected in ths field, none have known associations and we identify them as extragalactic sources.

\subsection{L673} L673, see Fig.~\ref{fig:figs1}\,(c), is a large star forming region in the Aquila Rift and the field observed here covers only part of it. Radio counterparts are detected for the known sub-mm cores L673~SMM-1 and L673~SMM-2 (Visser et~al. 2002). A third sub-mm core, identified as L673~SMM-7 in Visser et~al. (2002), which lies to the north-west of L673~SMM-1 is not detected. This core has no detected molecular outflow (Visser et~al. 2002) and is assumed starless. The two radio counterparts detected in the AMI-LA maps have been denoted L673~SMM-1 and L673~SMM-2 in Table~\ref{tab:fluxes}. However we note that although L673~SMM-1 is coincident with the sub-mm core detected in the SCUBA data of Visser et~al. (2002), the radio counterpart to L673~SMM-2 is offset slightly to the north of the sub-mm position. The peak flux density of the L673~SMM-1 radio counterpart is also coincident with the candidate embedded object [DCE08]-027, which is a Group 1 candidate according to the classification of Dunham et~al. (2008). A second candidate [DCE08]-028 (Group 5) is offset very slightly to the north-east, away from the peak. Although the sub-mm core L673~SMM-2 is very close in position to the candidate embedded object [DCE08]-029 (Group 1) the slight offset of the radio emission to the north makes the radio counterpart more coincident with [DCE08]-030 (Group 3). However the two candidate sources are separated by less than one synthesized beam and it is difficult to correctly assign the radio emission to a single candidate with observations of this resolution.

Aside from the previously mentioned sources there is one further object in this field, which has no known radio or protostellar counterpart and therefore we ascribe to being an extragalactic source.

\subsection{CB188} CB188, see Fig.~\ref{fig:figs1}\,(d), shows an elongated radio structure with a slight localized enhancement at the position of the sub-mm core (CB188~SMM) as detected by the Max Planck Millimetre Bolometer (MAMBO; Kauffmann et~al. 2008) and SHARC-II (Wu et~al. 2007). CB188~SMM-1 is a known Class I source (Launhardt et~al. 2010) and has molecular outflows which were measured by Yun \& Clemens (1994). The radio emission seen at 16\,GHz by the AMI-LA shows the same spur of emission extending to the north away from the position of the sub-mm core, which is seen at 450, 850 and 1300\,$\mu$m (Launhardt et~al. 2010). However the bulk of the extended radio emission seen in the AMI-LA map of this object is extended East--West, whereas the molecular outflows associated with CB188~SMM-1 are seen to be almost exactly coincident with the core and therefore are thought to be ``pole-on'', and to extend along the line of sight towards this object (Yun \& Clemens 1994). It is consequently difficult to associate the localized enhancements of radio emission along this extension as being associated with the outflows. Given the unusual extended nature of this emission compared with the other fields in our sample we are hesitant to associate it entirely with CB188~SMM. Taking this into account, the complex radio morphology of this object prevents us from extracting an integrated flux density for CB188~SMM-1 alone, and so instead we place a lower limit on the integrated radio flux density, corresponding to the peak flux density at the position of the core. 

\subsection{L673-7} 
\begin{figure*}
\centerline{\includegraphics[width=0.4\textwidth]{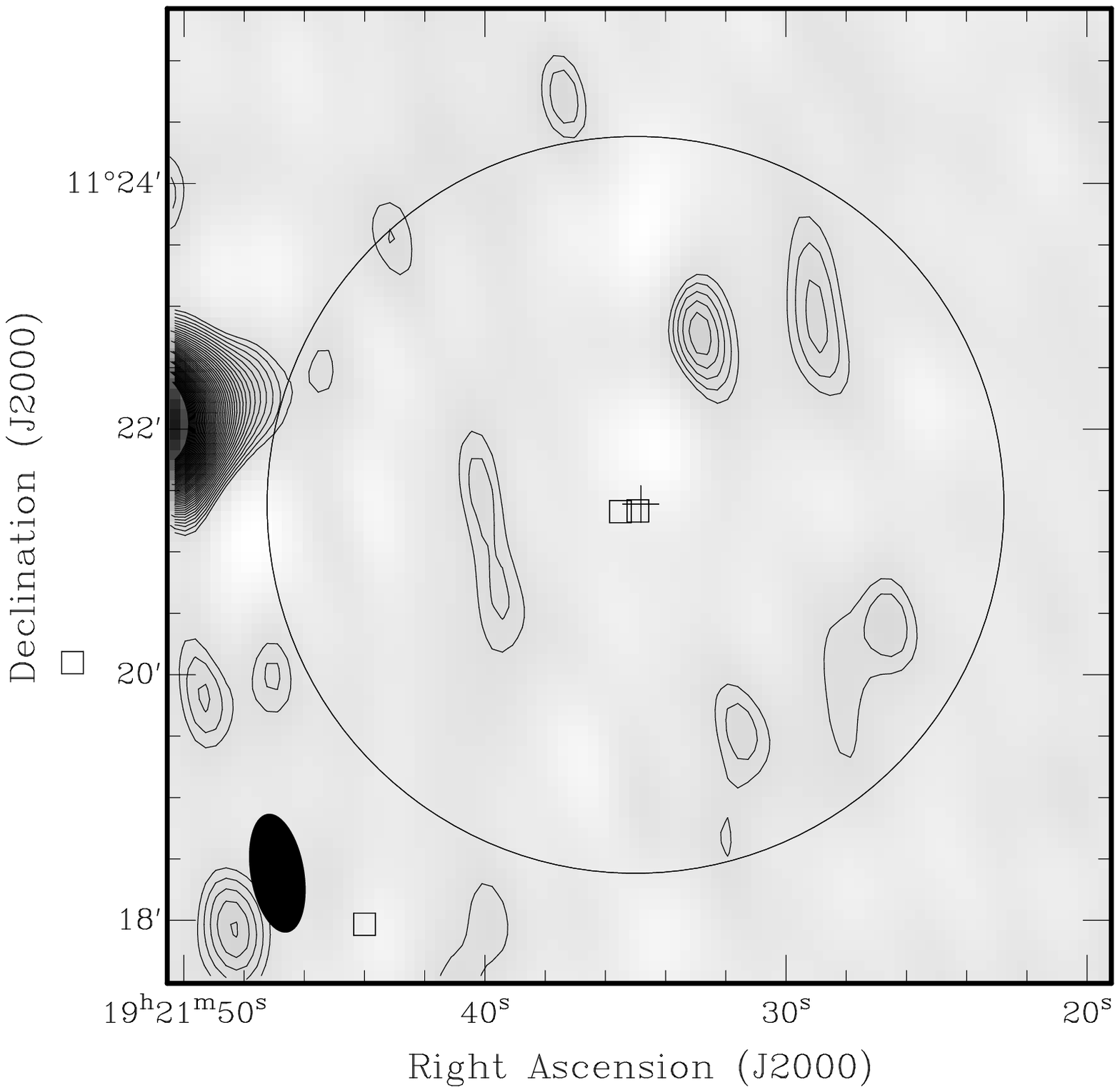}\qquad\includegraphics[width=0.4\textwidth]{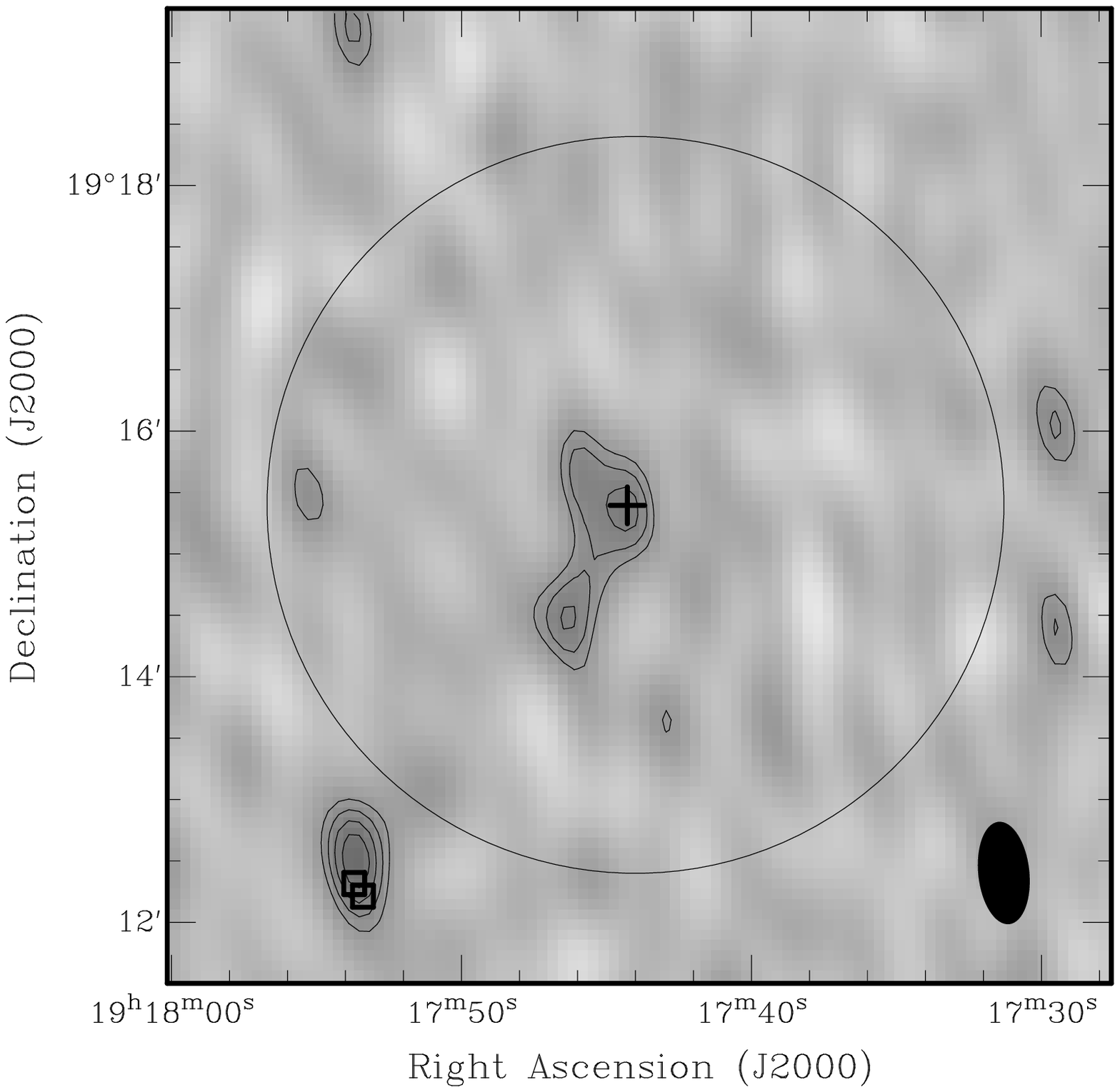}}
\centerline{(a)\hspace{0.4\textwidth}(b)}
\centerline{\includegraphics[width=0.4\textwidth]{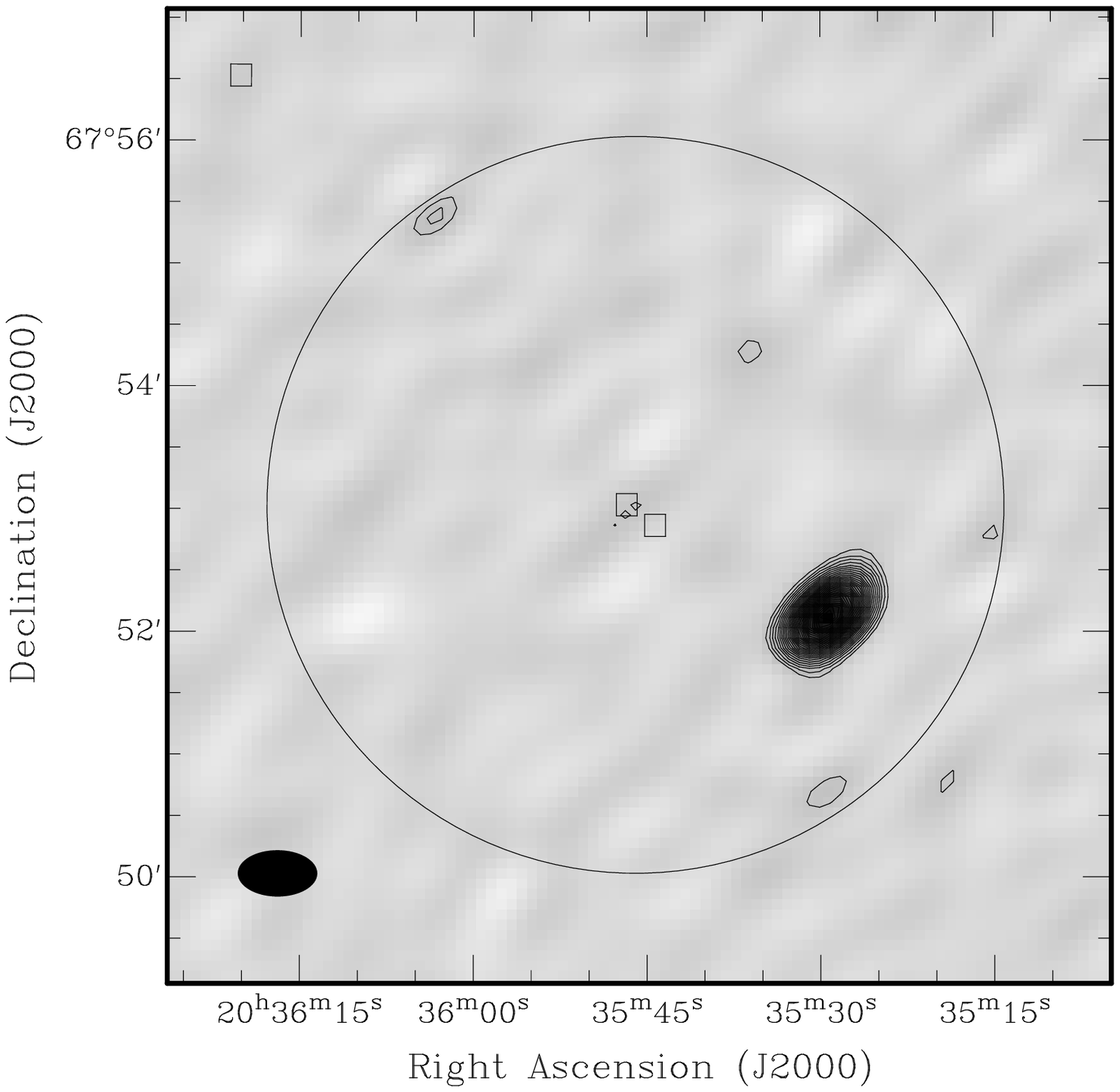}\qquad\includegraphics[width=0.42\textwidth]{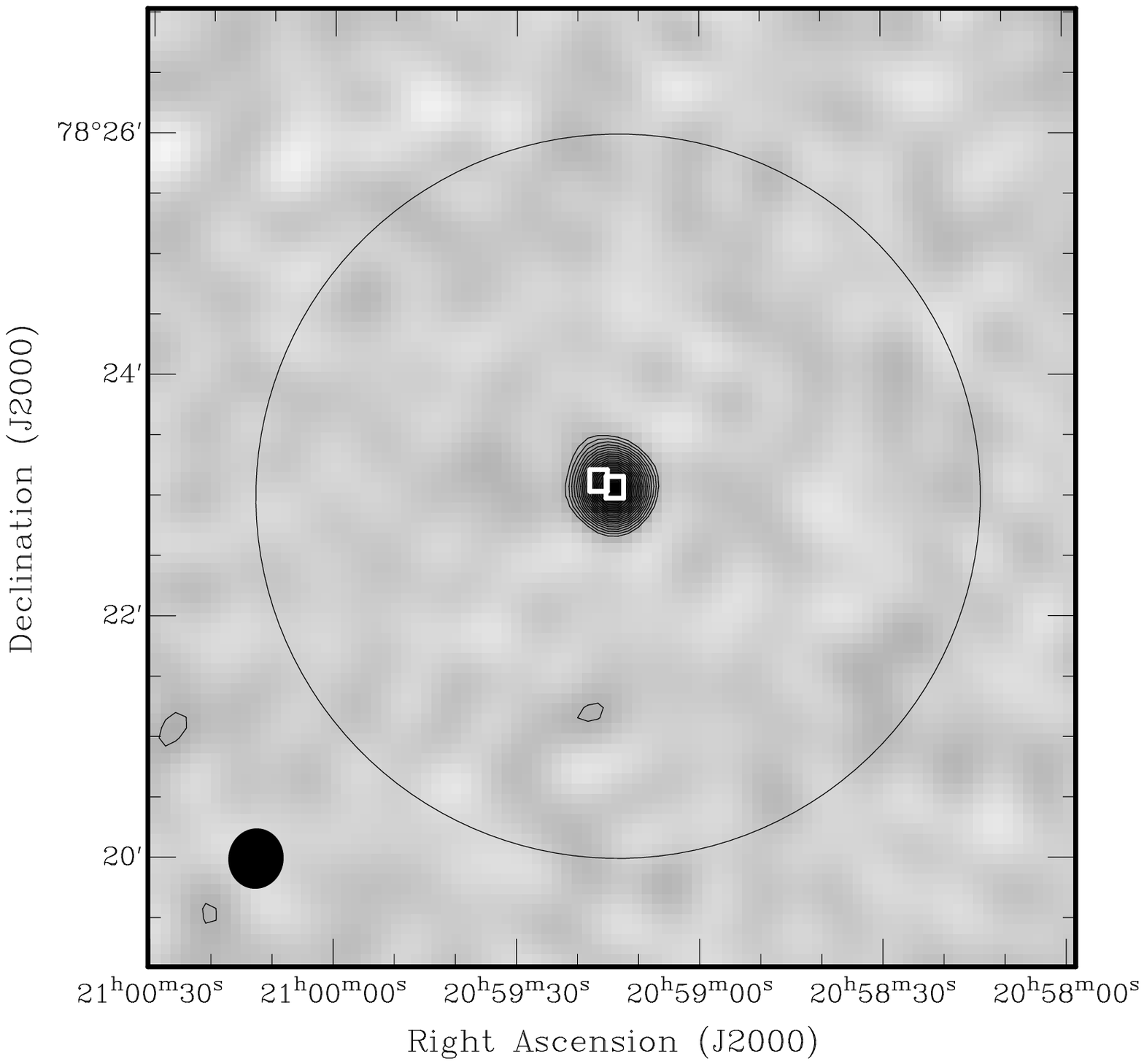}}
\centerline{(c)\hspace{0.4\textwidth}(d)}
\caption{(a) L673-7; (b) L723; (c) L1152; and (d) BERN~48. AMI-LA data are shown as greyscale and contours linearly from 3\,$\sigma$, where $\sigma=19, 22, 49$ and $23\,\mu$Jy\,beam$^{-1}$.  \emph{Spitzer} candidate embedded objects from Dunham et~al. (2008) are indicated by crosses, and the positions of sub-mm cores are shown as unfilled squares, see text for details. The AMI-LA primary beam FWHM is shown as a circle and the AMI-LA synthesized beam is shown as a filled ellipse in the lower left corner. \label{fig:figs2}}
\end{figure*}

Although L673-7-IRS, see Fig.~\ref{fig:figs2}\,(a), has previously been assumed to be a starless core, Dunham et~al. (2008) classified it as a Group 1 protostellar candidate based on its \emph{Spitzer} data, albeit a low luminosity $L_{\rm{IR}}=0.017$\,L$_{\odot}$ one. In spite of this we detect no source at the position of L673-7-IRS ([DCE08]-031). The two sources which are detected have no known associations and we tentatively identify them as extragalactic.

\subsection{L723} L723~SMM-1, see Fig.~\ref{fig:figs2}\,(b), is a known Class O object with a very powerful molecular outflow (Bontemps et~al. 1996). As its bolometric luminosity is also quite low, this places it in the early stages of Class O evolution. We detect L723~SMM-1 just outside the AMI-LA primary beam FWHM coincident with its known sub-mm position. The AMI-LA field is centred on [DCE08]-026, which we shall denote L723-IRS, and which is classified as a Group 6 embedded protostellar candidate by Dunham et~al. (2008). Group 6 candidates are known to be \emph{not} associated with regions of high volume density and are therefore considered to be unlikely candidates, with the caveat that most of the dust continuum surveys used to determine regions of high volume density are limited to $M\geq 0.1-1.0$\,M$_{\odot}$. At 16\,GHz we see a clear radio association to L723-IRS, with a second object offset to the south which may also be associated.

\subsection{L1152} We detect no source at the position of L1152-IRS, see Fig.~\ref{fig:figs2}\,(c). Only one source is visible within the AMI-LA primary beam which we identify with NVSS~203529+675207. This source has a primary beam corrected flux density of $S_{16.}=2.47\pm0.13$\,mJy, which implies a spectral index of $\alpha_{1.4}^{16}=0.15\pm0.03$. Due to its position offset from L1152-IRS we do not associate it with the core. However, we note that observations in the direction of L1152-IRS with the AMI-LA are heavily affected by radio interference and the noise level of this observation is high compared to the rest of the sample.

\subsection{L1148} 
\begin{figure}
\centerline{\includegraphics[width=0.5\textwidth]{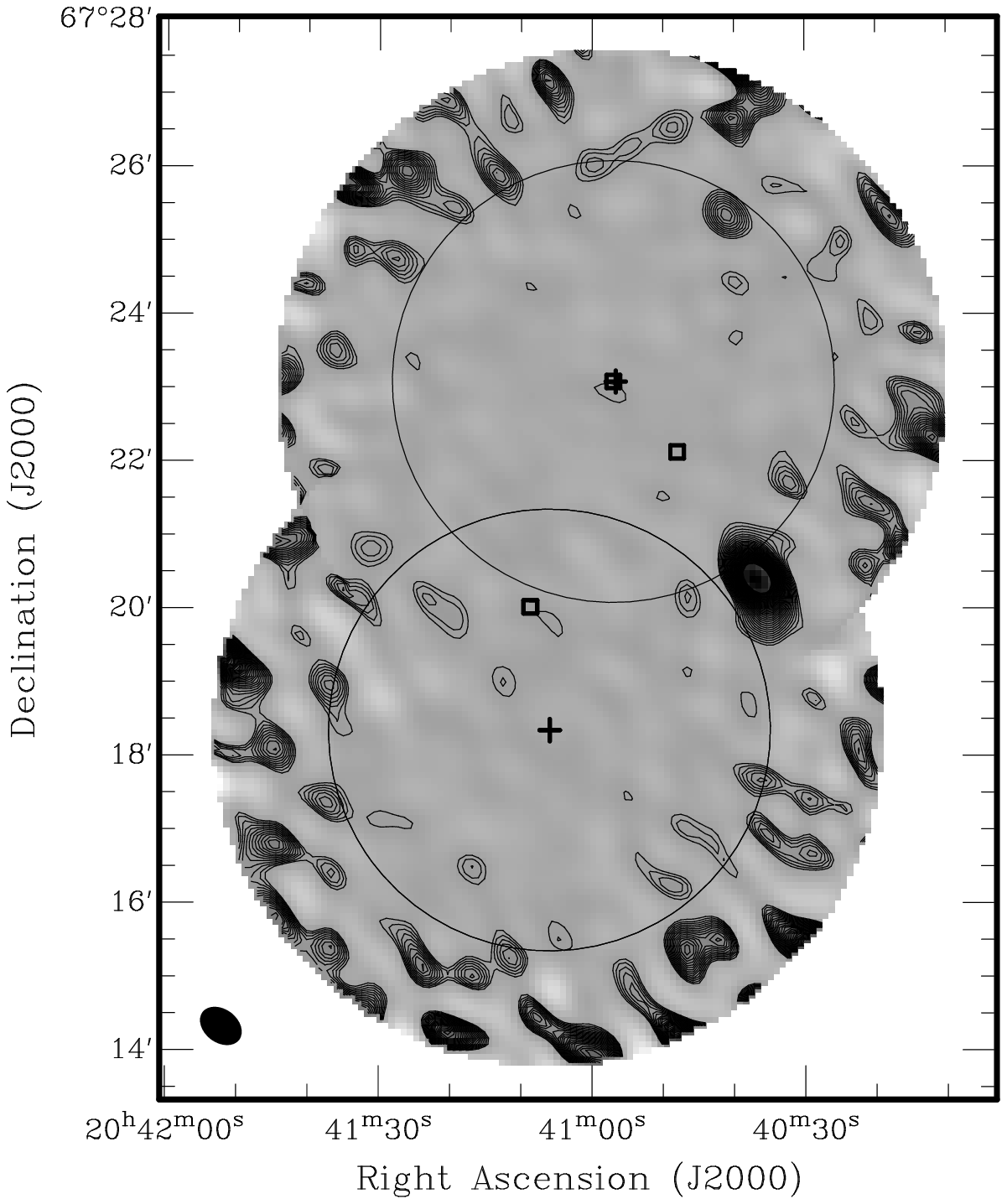}}
\caption{ Primary beam corrected mosaic of L1148. AMI-LA data are shown as greyscale and contours linearly from 3\,$\sigma$, where $\sigma=22\,\mu$Jy\,beam$^{-1}$.  \emph{Spitzer} candidate embedded objects from Dunham et~al. (2008) are indicated by crosses, and the positions of sub-mm cores are shown as unfilled squares, see text for details. The AMI-LA primary beams are shown as circles and the AMI-LA synthesized beam is shown as a filled ellipse in the lower left corner. \label{fig:l1148}}
\end{figure}

The L1148 cloud appears as two filaments of emission in the MAMBO 1.2\,mm dust maps of this region (Kauffmann et~al. 2008).  We make a weak detection $(3\,\sigma)$ of some extended radio emission coincident with the northern most filament, see Fig.~\ref{fig:l1148}, also referred to as L1148B in SHARC-II observations of this region. This northern filament is also coincident with a bright \emph{Spitzer} source at 24 and 70\,$\mu$m, L1148-IRS, which has been proposed as a VeLLO (Kauffmann et~al. 2005). Unlike other VeLLOs, such as L1014-IRS and IRAM~04191-IRS, no molecular outflow has yet been detected from L1148-IRS and its status as a sub-stellar mass Class O source is still debatable. Dunham et~al. (2008) place it in Group 3 of their classification scheme, making it a \emph{probable} protostar. If the enhancement of radio emission seen by AMI-LA is associated with ionization from a previously undetected molecular outflow or stellar wind from L1148-IRS this would support the case for its classification as a VeLLO. The second of the \emph{Spitzer} protostar candidates for the L1148 cloud, classified as Group 6, shows no corresponding radio emission in the AMI-LA data. With no known association, we identify the one further source detected in this field as extragalactic.

\subsection{Bern~48} Bern~48 (also known as RNO129 and HH198) is a borderline Class I/II object (Connelley et~al. 2008; Wu et~al. 2008). The AMI-LA observations towards Bern~48, see Fig.~\ref{fig:figs2}\,(d), show a slightly resolved compact source ($\Omega_{\rm{src}}\approx 5''$) with a rising spectrum. The source is clearly extended in the sub-mm and on larger scales displays a cometary morphology in the 1.2\,mm MAMBO maps (Kauffmann et~al. 2008). The AMI-LA measures a flux density of $S_{16}=615\pm39\,\mu$Jy, with a spectral index measured across the AMI band, although poorly constrained, of $\alpha_{14.9}^{17.9}=1.25\pm0.96$. A spectral index of $\simeq 1$ is consistent with that seen towards hypercompact {\sc Hii} regions, or indeed which may be expected from a stellar wind. If Bern~48 is a {\sc HCHii} region of consistent density this would imply an emission measure of $>10^9$\,pc\,cm$^{-6}$ and a mass of $\geq 100$\,M$_{\odot}$. Mass estimates for Bern~48, derived from MAMBO flux densities (Kauffmann et~al. 2008) find $M_{\rm{tot}}=2.06\pm0.02$\,M$_{\odot}$ inconsistent with this prediction. 

\subsection{L1014} 
\begin{figure*}
\centerline{\includegraphics[width=0.4\textwidth]{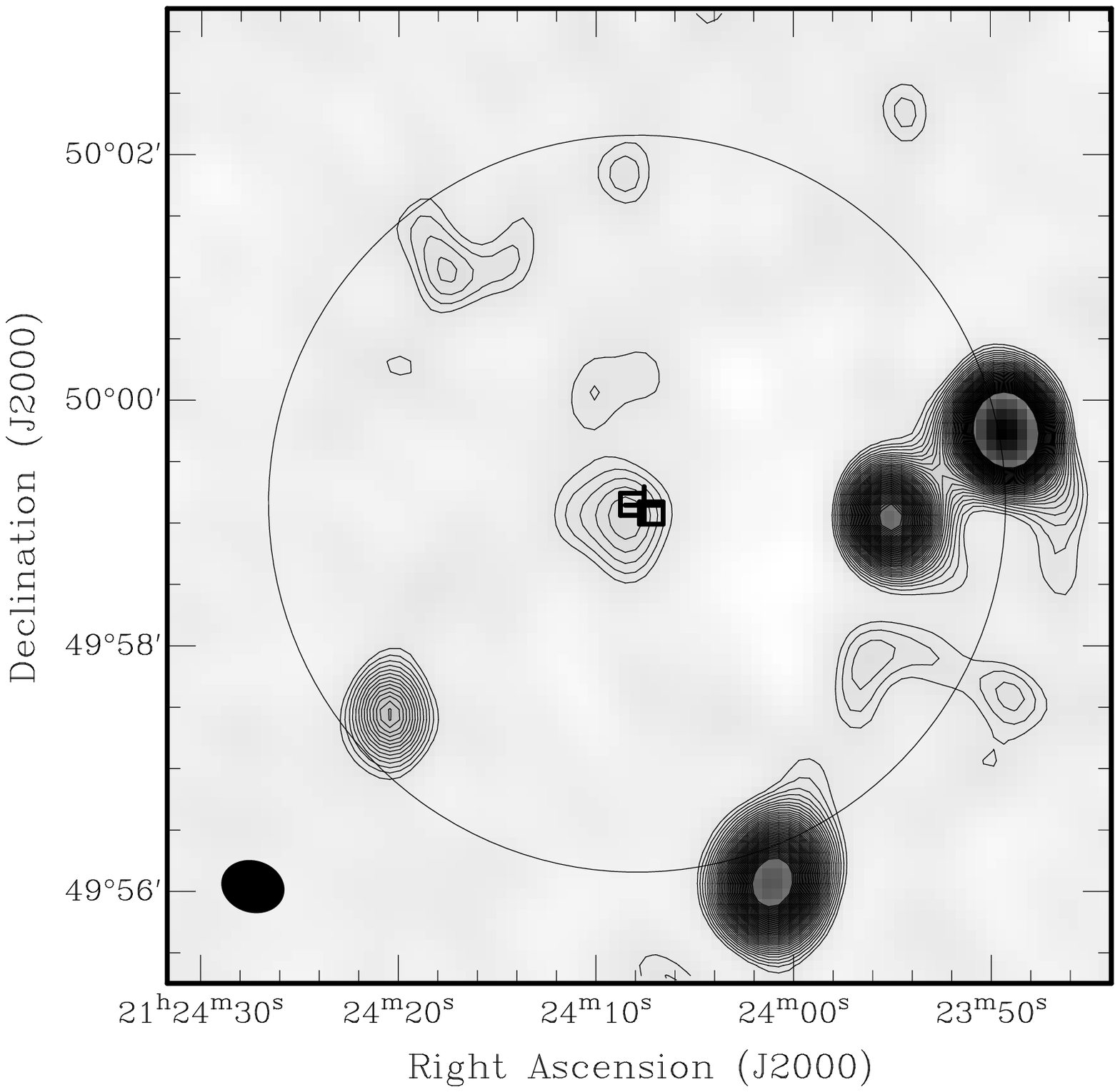}\qquad\includegraphics[width=0.4\textwidth]{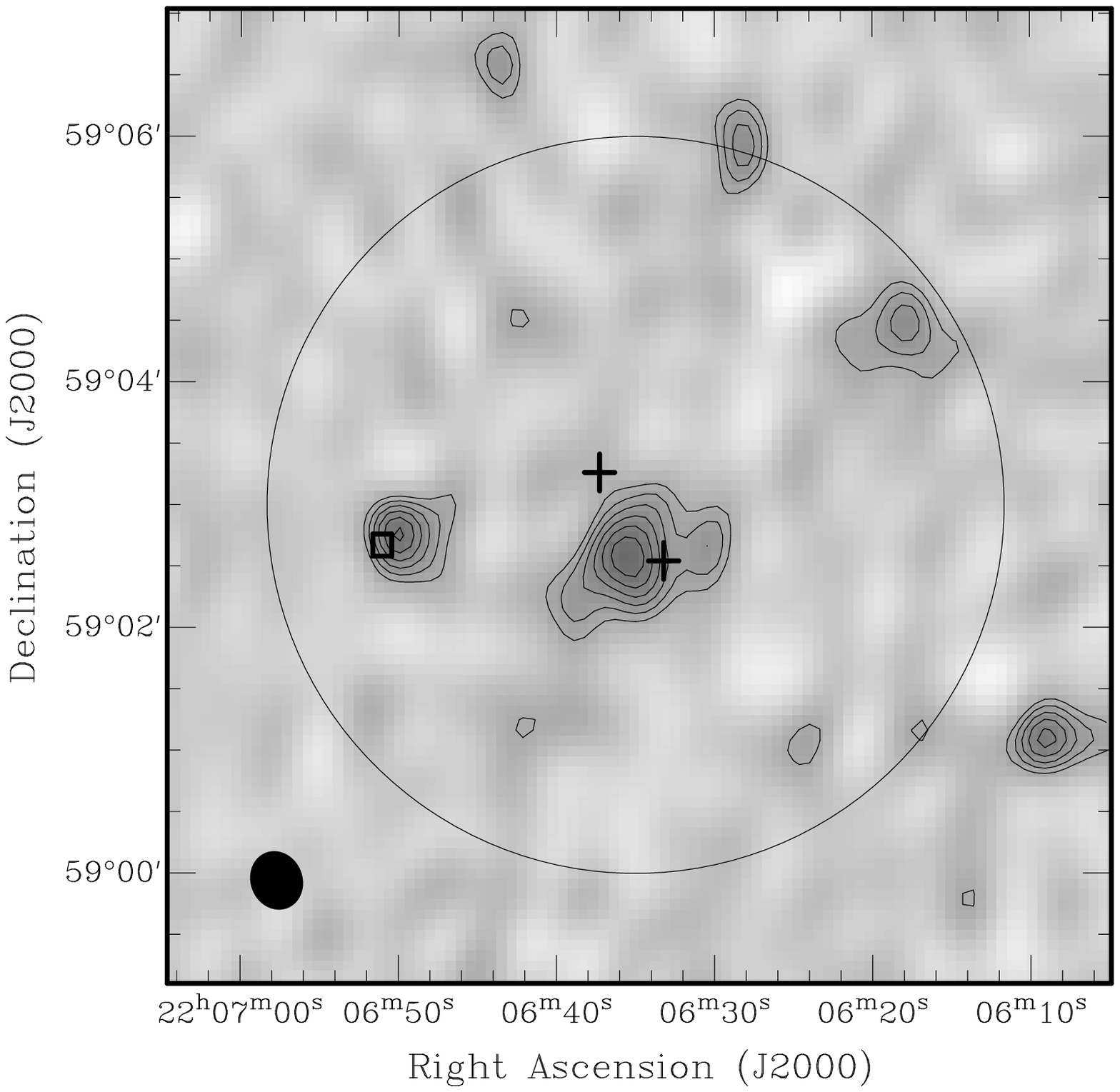}}
\centerline{(a)\hspace{0.4\textwidth}(b)}
\centerline{\includegraphics[width=0.4\textwidth]{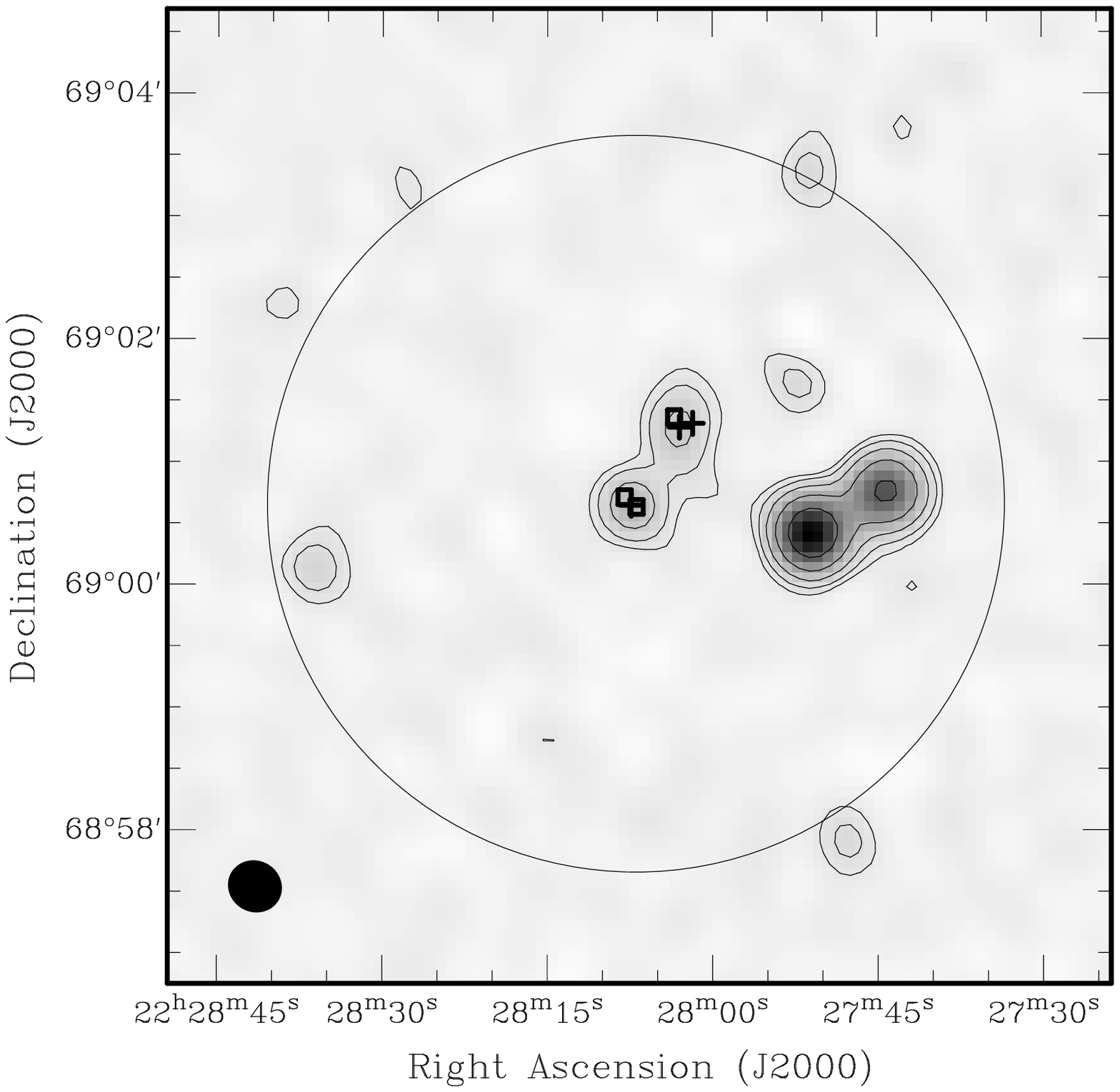}}
\centerline{(c)}
\caption{(a) L1014; (b) L1165; and (c) L1221. AMI-LA data are shown as greyscale and contours. In (a) \& (b) contours are linear from 3\,$\sigma$, where $\sigma=21\,\mu$Jy\,beam$^{-1}$ in both maps. In (c) contours are at 3, 6, 12, 24\,$\sigma$, where $\sigma=19\,\mu$Jy\,beam$^{-1}$. \emph{Spitzer} candidate embedded objects from Dunham et~al. (2008) are indicated by crosses, and the positions of sub-mm cores are shown as unfilled squares, see text for details. The AMI-LA primary beam FWHM is shown as a circle and the AMI-LA synthesized beam is shown as a filled ellipse in the lower left corner of each map.   \label{fig:figs3}}
\end{figure*}

The sub-mm core in L1014 was revealed not to be starless, as previously thought (Visser et~al. 2002), by the \emph{Spitzer} c2d programme and became the first VeLLO object (L1014-IRS; $L_{\rm{IR}}\simeq 0.09$\,L$_{\odot}$; Young et~al. 2004). Deep VLA observations at 6 and 3.6\,cm revealed a radio counterpart with a rising spectral index of $\alpha_{4.86}^{8.46}=0.37\pm0.34$. The AMI-LA measures a flux density of $S_{16}=299\pm27\,\mu$Jy towards L1014-IRS, constraining the spectral index to be $\alpha_{4.86}^{16}=0.71\pm0.11$. We observe the emission from L1014-IRS to be slightly extended, see Fig.~\ref{fig:figs3}\,(a), suggesting once more that the radio emission arises not from the central engine itself but from the more extended envelope or associated outflow.

There are also a number of bright radio sources in this field which can be found in the 1.4\,GHz NVSS catalogue and that we identify as steep spectrum radio sources. The two sources detected in this field which do not have an NVSS counterpart we also choose to identify as extragalactic as they have no known protostellar associations.

\subsection{L1165} L1165-IRS ([DCE08]-039) is classified as a Group 6 source by Dunham et~al. (2008), making it an unlikely candidate for an embedded object. In spite of this we observe an extended region of radio emission with a double-lobe structure, aligned east--west. Although the peak of neither lobe is exactly coincident with L1165-IRS we cautiously associate the larger lobe, which is nearer to the FIR source with this object. A second potential Group 6 core in L1165 ([DCE08]-040), with very low IR-luminosity, $L_{\rm{IR}}=0.008$\,L$_{\odot}$, is listed by Dunham et~al. (2008) to the north of L1165-IRS and we see no emission towards this object. The two detected sources which we do not associate with L1165-IRS do have any known radio or protostellar counterparts and we therefore identify them as extragalactic sources.

L1165~SMM-1, to the east of L1165-IRS, was identified as a Class I object by Visser et~al. (2002). The AMI-LA measurements of this source, see Fig.~\ref{fig:figs3}\,(b), show a relatively compact, isolated source slightly extended in the north--south direction. The spectrum of L1165~SMM-1 across the AMI band is relatively flat, suggesting that we seeing optically thin free--free emission. 

\subsection{L1221} The two sub-mm cores of the L1221 cloud (Young et~al. 2006; Wu et~al. 2007) were further resolved into three infra-red sources by \emph{Spitzer} (Young et~al. 2009). Centimetre radio observations of L1221 (Rodr{\'i}guez \& Reipruth 1998; Young et~al. 2009) at 3.6 and 6\,cm revealed what appears to be optically thin free--free emission from L1221-IRS3 (SMM-1) ($\alpha=-0.16\pm0.54$), but no detection in the direction of L1221-IRS1 and L1221-IRS2 (SMM-2). The AMI-LA 16\,GHz observations, see Fig.~\ref{fig:figs3}\,(c), clearly detect sources at both the location of L1221-IRS3 and towards L1221-IRS1 and L1221-IRS2 which are unresolved by the AMI-LA synthesized beam. The peak of the 16\,GHz emission towards L1221~SMM-2 is more closely associated with L1221-IRS1, which is classified as a Class I object, than L1221-IRS2 which is thought to be a YSOc. The flux density measured from the AMI-LA observations for L1221-IRS3 is $S_{16}=359\pm28\,\mu$Jy, in excess of a flux density extrapolated from the free--free model of the centimetre measurements by a factor of almost two. The 3.3\,mm measurements of Lee \& Ho (2005) indicate that the vibrational dust spectrum of L1221-IRS3 does not possess a flattened tail, such as might be expected in the presence of a protoplanetary disc that could explain the excess in terms of dust emission. If, as suggested by Young et~al. (2006), the centimetre emission observed towards L1221-IRS3 is due to shock ionization of the molecular outflow impacting on overdensities in the surrounding envelope (Curiel et~al. 1987; 1989; Anglada 1995; Shang 2004), then it is possible that the extra flux density at 16\,GHz is either indicative of a second shock component with a larger optical depth or may represent a steepening of the radio spectrum due to thermal emission from the immediate envelope of the central source as predicted by the model of Konigl (1984).
The two bright sources to the west of L1221 are coincident with a known radio source in the NVSS catalogue, the resolution of which does not resolve the two component sources, and we therefore identify them definitely as extragalactic objects. the two further sources detected in this field have no lower frequency radio counterpart, but also no known protostellar associations.

\section{Discussion}
\label{sec:disc}

\subsection{The nature of the cm-wave radio emission: the thermal dust contribution}

\begin{table*}
\caption{Physical properties of the AMI LA Protostar sample. \label{tab:phys}}
\begin{tabular}{lccccccccc}
\hline\hline
Name & $D$    & $L_{\rm{IR}}$ & $L_{\rm{bol}}$ & refs. & $F_{\rm{out}}$ & refs. & $S_{16}^{\rm{pred}}$ & $\log{S_{1.8{\rm{cm}}}D^2}$  & Class \\
     & (pc) & (L$_{\odot}$) & (L$_{\odot}$) & & ($10^{-5}$\,M$_{\odot}$\,km\,s$^{-1}$\,yr$^{-1}$) && ($\mu$Jy) & (mJy\,kpc$^2$) & \\
\hline
IRAM~04191 & 140$\pm$10 & 0.023 & 0.15, 0.13$\pm$0.03,0.12$\pm$0.02  & 1,2,10 & 1.5,2 & 3 & 45 & $-2.470$ & 0,VeLLO\\
IRAS~04191 & 140$\pm$10 & 0.25  & 0.3,0.34,0.5,0.64 & 2,8,9,10 & - & - &34 & $-3.000^{\ast}$ & I \\
L1521F-IRS	  & 140$\pm$10 & 0.015 & 0.13$\pm$0.02 & 10 &  -  & - & 24 & $-2.550$ &0,VeLLO\\
B35A~SMM-1      & 400$\pm$40 & 0.141 & 14,15 & 1,4 & 0.9$^c$ & 12 & 45 & $-0.821$ & I\\
L673~SMM-1 & 300$\pm$100& 0.201 & 2.8,1.0 & 1,4 & 0.6 & 4 & 5 & $-1.388$ & I\\
L673~SMM-2 & 300$\pm$100& 0.138 & 2.8,0.4 & 1,4 & 4.3,2.6 & 15,4 & 12 & $-1.584$ & 0\\
CB188~SMM-1   & 300        &   -    & $1.94\pm0.2$,1.5 & 1,5 & 3.5$^c$ & 5 & 18 &$-1.795^{\dagger}$& I\\
L673-7-IRS    & 300$\pm$100& 0.017 & 0.09$\pm$0.03 & 10 & - & - & 21 & $-2.290^{\ast}$ & 0,VeLLO\\
L723-IRS      & 300$\pm$100& 0.048 &  -   & - & - & - & - & $-1.703$  & 0,VeLLO\\
L723~SMM-1 & 300$\pm$100&   -    & 2,3   & 1 & 37 & 14 & 69 & $-1.253$  & 0\\
L1152~SMM-1 & 325        &   -    & 3.3 & 1 & 0.47 & 14 & 15 & $-1.809^{\ast}$& 0/I\\
L1148-IRS1     & 325$\pm$25 & 0.081 &$0.23\pm0.17$,0.15& 10 & - & - & - & $-2.120$ & 0,VeLLO\\
L1148-IRS2     & 325$\pm$25 & 0.003 & - & - & - & - & - & $-1.994^{\ast}$ & 0,VeLLO\\
Bern~48	  & 200        &   -    & $11.05\pm0.20$,13.5 & 6 & - & - & 39 & $-1.609$ & I/II\\
L1014-IRS     & 250$\pm$50 & 0.087 & 0.3,0.34$\pm$0.11 & 10,11 & 0.004$^c$ &7& 6 & $-1.729$& 0,VeLLO\\
L1165-IRS & 300$\pm$50 & 0.030 & - & - & - & - & - & $-2.032$ & 0,VeLLO\\
L1165~SMM-1  & 300$\pm$50 &  -     & 11.7 & 4 & 1.1 & 4 & 33 & $-1.663$ & I\\
L1221-IRS1& 250$\pm$50& 1.12   & 2.5  & 10 & - & -& 32 & $-1.649$& I\\
L1221-IRS3& 250$\pm$50& 0.36   & $0.93\pm0.16$  & 10 & 10.1$^c$ & 13 & 29 & $-1.649$& 0\\
\hline
\end{tabular}
\begin{minipage}{\textwidth}{
$^{\ast}$ $3\,\sigma$ upper limit.\\
$^{\dagger}$ lower limit, see text for details.\\
$^c$ Value of $F_{\rm{out}}$ from the literature corrected for $D$.\\
(1) Furuya et~al. 2003; 
(2) Dunham et~al. 2006; 
(3) Andr{\'e} et~al. 1999; 
(4) Visser et~al. 2002; 
(5) Yun \& Clemens 1994; 
(6) Connelley et~al. 2008; 
(7) Bourke et~al. 2005;
(8) Hayashi et~al. 1994;
(9) Ohashi et~al. 1996;
(10) Dunham et~al. 2008;
(11) Young et~al. 2004;
(12) Myers et~al. 1988;
(13) Umemoto et~al. 1991;
(14) Bontemps et~al. (1996);
(15) Armstrong \& Winnewisser (1989).
}
\end{minipage}
\end{table*}

At 16\,GHz we must be aware of the potential contribution to our measured flux densities from the vibrational, or Planckian, dust spectrum. Where available we use sub-mm flux densities from the literature at wavelengths from 350 to 1300\,$\mu$m to constrain the thermal dust emission at 1.8\,cm. In general we use flux densities extracted from $40''$ apertures, with notable exceptions being those objects measured by Visser et~al (2002) who used a $50''$ aperture, and IRAM~04191-IRS for which $60''$ apertures are assumed in order to accommodate the resolution of the 1.3\,mm data (Dunham et~al. 2006). We assume optically thin dust emission with $\beta=1.5$ and fit SEDs with a single-temperature greybody spectrum:
\begin{equation}
S_{16}^{\rm{pred}} \propto \nu^{\beta} B_{\nu}(T_{\rm{dust}}).
\end{equation}

The flux densities predicted from these fits are listed in Column [8] of Table~\ref{tab:phys}, which also lists the physical parameters of the cores taken from the literature. The predicted values of the vibrational dust emission are in general low compared to the rms noise in the datasets, detectable at 3\,$\sigma$ only in a few fields.

 \subsection{The nature of the cm-wave radio emission: free--free emission}
\label{sec:fcorr}

Massive star formation regions have been known for a long time to produce radio emission, however the low luminosity protostars in this sample do not produce a sufficient ionizing flux to maintain a photoionized {\sc Hii} region such as those found in hypercompact or ultracompact {\sc Hii}. Instead it is likely that an excess of radio flux density relative to the thermal dust emission in these objects is due to either a shocked molecular outflow or a stellar wind. A completely ionized stellar wind (Panagia \& Felli 1975) was shown to be feasible by Curiel et~al. (1989) only for objects $\ll 1\arcsec$ in size, as it would produce radio flux densities far in excess of those measured, e.g. at 16\,GHz for a wind of $5''$ angular size and assuming an electron temperature $T_{\rm{e}}=10^4$\,K, similar to Bern~48, we would observe a flux density of $S_{16}\simeq50$\,Jy. Since the majority of our observed sources are extended on scales of greater than an arcsec we find this possibility untenable. The alternatives are that we are seeing radio emission from a neutral wind or molecular outflow, which is being shock ionized as it impacts on overdensities within the surrounding interstellar medium or the infalling cold dynamically collapsing medium (Curiel et~al. 1987), or we are seeing radio emission from a partially ionized and possibly collimated flow (Reynolds 1986).

The cm-wave flux density from regions of shock ionization has been shown to be (Curiel et~al. 1987; 1989)

\begin{eqnarray}
\nonumber \left(\frac{S_{\nu}}{\rm{mJy}}\right) & = & 3.98\times10^{-2} \left(\frac{\dot{M}_{\ast}}{10^{-7}\rm{M}_{\odot}\rm{yr}^{-1}}\right) \left(\frac{V_{\ast}}{100\rm{km\,s}^{-1}}\right)^{0.68} \\
 && \times \left(\frac{D}{\rm{kpc}}\right)^{-2}\left(\frac{T}{10^4\,\rm{K}}\right)^{0.45}\left(\frac{\Omega}{4\pi}\right)\left(\frac{\nu}{5\,\rm{GHz}}\right)^{-0.1}.
\end{eqnarray}

\noindent
Following Anglada et~al. (1996; 1998) we may re-express this relation at 16\,GHz as
\begin{equation}
\left(\frac{F_{\rm{out}}}{\rm{M}_{\odot}\,\rm{yr}^{-1}\,\rm{km\,s}^{-1}}\right) = \frac{2.97\times10^{-4}}{\eta} \xi(\tau) \left(\frac{S_{\nu}D^2}{\rm{mJy\,kpc}^2}\right),
\end{equation}
where $\eta=(\Omega/4\pi)$ is the fraction of the stellar wind being shocked, $F_{\rm{out}}$ is the outflow force or momentum flux, equivalent to the rate of outflow momentum and often computed as $F_{\rm{out}} = (P/\tau_{\rm{dyn}})$. We also introduce $\xi(\tau)=\tau/(1-\rm{e}^{-\tau})$ to allow for the fact that the radio emission may not be optically thin (Anglada et~al. 1998), although this dependence on the optical depth, $\tau$, is very weak. 

The parameter $\eta$ may be considered as the efficiency of the shock and, following Anglada (1995), a value of $\eta=1$ allows us estimate the minimum outflow force required to explain the cm-wave emission as shock ionization. 
\begin{figure}
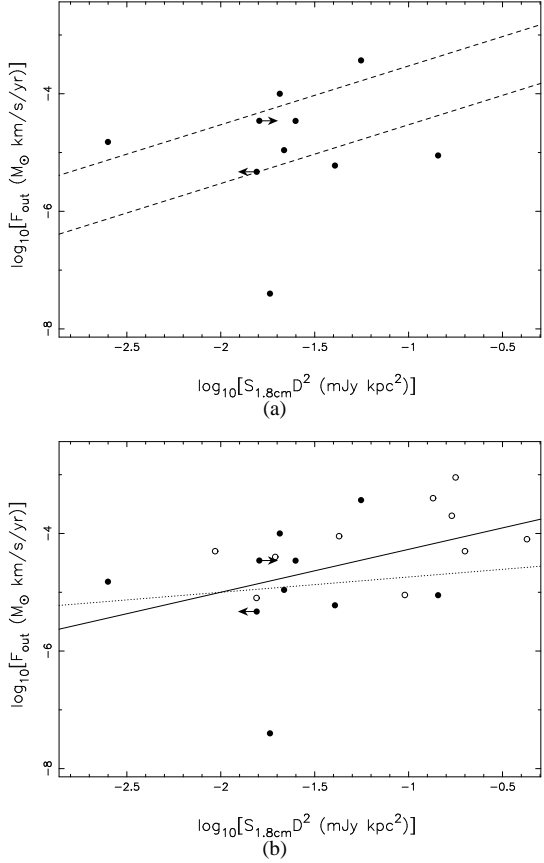

\centerline{\includegraphics[angle=-90,width=0.4\textwidth]{./corrFOUT.ps}}
\centerline{(a)}
\centerline{\includegraphics[angle=-90,width=0.4\textwidth]{./corrFOUText.ps}}
\centerline{(b)}
\caption{Distribution of radio luminosity vs. outflow force, see Table~\ref{tab:phys} for references. (a) Data points are radio luminosities derived from the AMI-LA measurements at 16\,GHz, see Table~\ref{tab:phys}. The dashed lines show the theoretical relationship between 1.8\,cm radio luminosity and outflow force (Curiel et~al. 1989) for an efficiency of $\eta=1$ (minimum required force; lower line) and an efficiency of $\eta=0.1$, see text for details. (b) Filled data points are radio luminosities derived from the AMI-LA measurements at 16\,GHz as before; unfilled data points are from Anglada et~al. (1998) extrapolated to 16\,GHz. The dotted line shows a linear regression to the AMI-LA data alone, and the solid line shows the regression to the combined data sets.
\label{fig:fcorr}}
\end{figure}
In Fig.~\ref{fig:fcorr}\,(a) we plot the correlation of the radio luminosity as measured at 1.8\,cm from this paper against the outflow force for each object where a value is available in the literature. For values and references see Table~\ref{tab:phys}. We also plot the predicted outflow force from the Curiel model for an efficiency of $\eta=1$ (minimum required force; lower line) and for an efficiency of $\eta=0.1$. 

There are considerable uncertainties in outflow force measurements and variations in how the force is estimated. Some estimates are derived from the total momentum observable in the outflow lobes, divided by a measure of the outflow dynamical time (e.g. B35A~SMM-1 or CB188). Others use a measure of the force derived within an annulus around the outflow origin (Bontemps et~al. 1996; e.g. L723~SMM-1, L1165~SMM-1), which provides a better estimate of the force for outflows which have been only partially mapped. Even analyses which use similar methods can have variations, e.g. whether the maximum velocity in an outflow is used to compute the dynamical time or some average/typical velocity. Furthermore, missing flux density (in interferometric studies), high optical depths of CO emission in the outflow line wings and inclination effects which reduce the observed outflow velocities all tend to reduce the momentum actually observed. 

Many studies follow the lead of Bontemps et~al. (1996) and increase their outflow forces by a factor of 10 to account for optical depth and inclination effects (e.g. Visser et~al. 2002). Other authors are able to correct for optical depth using observations of the rarer isotopologues of CO (e.g. L1221-IRS3; Umemoto et~al. 1993) or note that the inclination corrections will be minimal (e.g. L1014-IRS; Bourke et~al. 2005). A consistent approach for the outflows in our sample is not possible. Instead we use the values as stated in the literature, corrected only if we assume a different distance ($F_{\rm{out}}\propto D$). For a large enough sample of protostars, the trends should still be apparent, albeit with a large scatter, but the force for any particular outflow could be under-estimated by a factor of $\approx 10$.

From Fig.~\ref{fig:fcorr}\,(b) we can see that three objects lie below the calculated minimum outflow force: L1014-IRS, L673~SMM-1 and B35A~SMM-1. Missing flux and optical depth effects are not taken into account in the outflow force of L1014-IRS (Bourke et~al. 2005). However, even if the value were scaled upwards by an order of magnitude it would still fall below the predicted minimum force limit. A similar upwards scaling for B35A~SMM-1 could be required, as optical depth and inclination were not accounted for, unlike L673~SMM-1, which has been scaled upwards following the prescription of Bontemps et~al. (1996). If all the objects were treated in the same way it is likely that only L1014-IRS and L673~SMM-1 would remain below the minimum line. Shirley et~al. (2007) found the same situation for L1014-IRS at 3.6 and 6\,cm and concluded that L1014 possessed an additional ionization mechanism aside from shock ionization. This additional component is most likely to be provided by a self-shocked spherical wind (Panagia \& Felli 1975; Wright \& Barlow 1975; Reynolds 1986). A caveat to this explanation is that spherical winds are usually associated with more evolved Class II objects. 

As discussed earlier, a fully ionized spherical wind would be likely to produce a flux density far in excess of that which we measure for L1014-IRS, particularly as the radio emission is slightly extended. An alternative is to consider the model of Reynolds (1986), which allows for partially ionized, collimated winds. An advantage of this model above that of a spherical wind is that it allows for values of the radio spectral index from an unresolved outflow of $\alpha > 0.6$, as has been observed in a number of cases. This model assumes power-law dependencies of physical parameters on the radial distance, $r$, from the central source. The first of these is the jet half-width, $w$, which is assumed to obey $w=w_0(r/r_0)^{\epsilon}$ where $r_0$ represents the ``ionization radius'' at which the thermal emission begins. For well collimated flows, $r_0$ and $w_0$ are related by the characteristic function $\theta_0=2w_0/r_0 \leq 0.5$. Reparameterizing the equations of Reynolds (1986) in terms of the physical parameters observed here we find that the predicted radio emission can be found using,
{\small
\begin{eqnarray}
\label{equ:s_rey}
\nonumber S_{\nu}(\rm{mJy}) = 3.2&\times&\theta_0 \left(\frac{r_0}{10^{15}\,\rm{cm}}\right)^2 \left(\frac{\nu}{10\,\rm{GHz}}\right)^{\alpha} \left(\frac{\nu_{\rm{m}}'}{10\,\rm{GHz}}\right)^{2-\alpha}\\
\nonumber &\times& \left(\frac{D}{\rm{kpc}}\right)^{-2}\left(\frac{T_{\rm{e}}}{10^4\,\rm{K}}\right) F_{\rm{R}}(q_{\tau},\alpha) x_0^{-2c} (\sin{i})^{1+c}\\
\end{eqnarray} 
}
where
{\small
\begin{eqnarray}
\label{equ:vm}
\nonumber\left(\frac{\nu_{\rm{m}}'}{10\,\rm{GHz}}\right)^{2.1} &=& 3.7\times 10^{-7} \theta_0^{-3} \left(\frac{F_{\rm{out}}}{10^{-5}\,\rm{M}_{\odot}\,\rm{yr}^{-1}\,\rm{km\,s}^{-1}}\right)^2\\ 
\nonumber & & \times \left(\frac{r_0}{10^{15}\,\rm{cm}}\right)^{-3} \left(\frac{V_{\ast}}{200\,\rm{km\,s}^{-1}}\right)^{-4}\left(\frac{T_{\rm{e}}}{10^4\,\rm{K}}\right)^{-1.35},\\
\end{eqnarray}
}
\begin{equation}
F_{\rm{R}}(q_{\tau},\alpha) = \frac{(2.1)^2}{q_{\tau}(\alpha-2)(\alpha+0.1)},
\end{equation}
and
\begin{equation}
c = (1+\epsilon+q_{\rm{T}})/q_{\tau} = \frac{\alpha-2}{2.1}.
\end{equation}

The parameters $q_{\rm{T}}$ and $q_{\tau}$, corresponding to the notation of Reynolds (1986), represent the power-law indices for temperature and optical depth, respectively. The values of these parameters for various models may be found in Table~1 of Reynolds (1986), as may the predicted value of $\alpha$. The $\sin{i}$ dependence in this formulation takes into account the inclination angle, $i$, of the outflow to the observer's line of sight. $x_0$ is the fractional ionization of the gas, $x$, at the ionization radius, which is a tunable parameter in the model and affects the optical depth through the free--free absorption coefficient , $\kappa_{\nu}$, in the standard way: $\tau\propto x^2$. A caveat to this model is that there is no explanation of how the ionization is produced, and a value must simply be assumed.

By combining Equations~\ref{equ:s_rey} and \ref{equ:vm}, we can see that this model predicts a correlation between the radio luminosity and the outflow force of,
\begin{equation}
F_{\rm{out}} \propto L_{\rm{rad}}^{2.1/2(2-\alpha)}.
\end{equation}
For instance, in the case where $\alpha=0.6$ this correlation becomes $F_{\rm{out}} \propto L_{\rm{rad}}^{0.75}$.

From the number of data points in the AMI-LA data alone which have measurements of $F_{\rm{out}}$, see Fig.~\ref{fig:fcorr}\,(a), there is no evidence of a correlation and the Pearson product momentum correlation coefficient is low: $r_{\rm{xy}}=0.12$. To enhance the trend we also include datapoints extrapolated from the measurements of Anglada et~al. (1998), Fig.~\ref{fig:fcorr}\,(b). These data are extrapolated from measurements at 3.6\,cm using a power law with spectral indices derived from data at 3.6 and 6\,cm. The flux density measurements, derived spectral indices and measured outflow forces for these data may be found in Table~5 of Anglada et~al. (1998). 

Including the additional data, the correlation between outflow force and radio luminosity is still weak ($r_{\rm{xy}}=0.40$). This may be due to a lack of data, but could also be a consequence of the uncertainties in the quoted values of $F_{\rm{out}}$. The correlation we find from the AMI-LA data alone is
{\small
\begin{eqnarray}
\nonumber \log[F_{\rm{out}} (\rm{M}_{\odot}\rm{yr}^{-1}\rm{km\,s}^{-1})] & = & -(4.48\pm1.26)\\
\nonumber &&+(0.26\pm0.74)\log[L_{\rm{1.8\,cm}} (\rm{mJy\,kpc}^2)].\\
\end{eqnarray}
}
If we also include the extrapolated data from Anglada et~al. (1998) we find
{\small
\begin{eqnarray}
\nonumber \log[F_{\rm{out}} (\rm{M}_{\odot}\rm{yr}^{-1}\rm{km\,s}^{-1})] & = & -(3.54\pm0.58)\\
\nonumber &&+(0.73\pm0.38)\log[L_{\rm{1.8\,cm}} (\rm{mJy\,kpc}^2)],\\
\end{eqnarray}
}
see Fig.~\ref{fig:fcorr}\,(b). This correlation is very similar to that found by Shirley et~al. (2007) at 3.6\,cm. 

Due to the short frequency coverage of the AMI-LA band and the low signal to noise of the objects in this sample we consider the errors on any spectral index derived from AMI-LA data alone too large to allow a reliable quantitative analysis. However, six of the objects in our sample have additional radio measurements at lower frequencies, these are listed in Table~\ref{tab:rad}, and we use these to examine the spectral behaviour of the radio emission from these cores. Where the source has not been detected we take a 3\,$\sigma$ upper limit on its flux density, with $\sigma$ equal to the rms noise as quoted in the reference. 
\begin{table*}
\caption{Radio Spectral Indices. \label{tab:rad}}
\begin{tabular}{lccccc}
\hline\hline
Name & $S(6\,\rm{cm})$ & $S(3.6\,\rm{cm})$ & $S(2(1.8)\,\rm{cm})$ & $\alpha$ & refs. \\
\hline
IRAM~04191 & $480\pm30$ & $140\pm20$ & $160\pm50$, $172.9\pm17$ & $0.45\pm0.20$ & 1, this work \\
L1521F     & -          & $<66$      & $141\pm16$ & $>1.10$ & 2, this work\\
B35A       & $<120$     &    -       & $943\pm67$ & $>1.77$ & 3, this work\\
L723-SMM1  & -          & $740\pm42$ & $690\pm29$ & $-0.10\pm0.21$ & 4, this work\\
L1014      & $88\pm11$  & $111\pm8$  & $299\pm21$ & $0.71\pm0.11$ & 5, this work\\
L1221-IRS3 & $192\pm27$ & $177\pm24$ & $359\pm19$ & $0.83\pm0.10$ & 6, this work\\
\hline
\end{tabular}
\begin{minipage}{\textwidth}{
[1] Andr{\'e} et~al. (1999);
[2] Harvey et~al. (2002);
[3] Rodr{\'i}guez et~al. (1989);
[4] Anglada et~al. (1991);
[5] Shirley et~al. (2007);
[6] Young et~al. (2009).
}
\end{minipage}
\end{table*}
We find that all of the sources have rising spectral indices, indicating that they all possess partially optically thick free--free emission, with the exception of L723~SMM-1. This source has a spectral index consistent with optically thin free--free emission. We also note that the outflow force quoted for L723~SMM-1 is an order of magnitude larger than that of any other source (although this incorporates the upwards scaling of Bontemps et~al. (1996)). Such behaviour may be consistent with the shock ionization taking place further from the central protostar, since the model of Curiel et~al. (1987; 1989) predicts that the distance of the shock from the star, $R$, is proportional to the outflow force,
\begin{equation}
\left(\frac{R}{\rm{AU}}\right)=1.22\times10^3 \tau^{-0.5} \left(\frac{F_{\rm{out}}}{\rm{M}_{\odot}\rm{yr}^{-1}\rm{km\,s}^{-1}}\right)^{0.5}
\label{equ:rad}
\end{equation}
(Anglada et~al. 1998). However, the more detailed treatment of Reynolds (1986) shows that although the optical depth is a power law function of the distance from the central star, $\tau(r)=\tau_0 (r/r_0)^{q_{\tau}}$, the index of the power-law can have a wide range of values, $-4 \leq q_{\tau} \leq -0.9$, depending on the morphology of the outflow and whether recombination takes place. Substituting these values into 
Equ.~\ref{equ:rad} we find
\begin{equation}
\left(\frac{R}{\rm{AU}}\right)\propto \left(\frac{F_{\rm{out}}}{\rm{M}_{\odot}\rm{yr}^{-1}\rm{km\,s}^{-1}}\right)^{-0.5\leq q_F \leq 0.77}.
\end{equation}
Given that this is an oversimplification, it is therefore perhaps unsurprising that we find no correlation between spectral index and outflow force ($r=0.12$). In the case of the standard spherical flow (Panagia \& Felli 1975; Wright \& Barlow 1975) $q_{\tau}=-3$, which would produce a correlation of $R\propto F_{\rm{out}}^{-1}$. We might then interpret the observed lack of correlation, both in this work and as found by Shirley et~al. (2007), as evidence that the majority of cm-wave emission from protostars is not produced, at least solely, via this model.

\subsection{The Very Low Luminosity Radio Correlation}
\label{sec:corr}

The correlation between radio and bolometric luminosity for lower mass protostars was first described for 3.6\,cm emission by Anglada (1995) and recently updated by Shirley et~al. (2007) who also included a correlation for 6\,cm emission. This correlation is thought to arise as a consequence of the correlation between outflow force and bolometric luminosity (Shirley et~al. 2007) as the degree of shock ionization is expected to be enhanced for more powerful outflows from higher luminosity sources (Anglada 1995). From our 1.8\,cm data we find correlations of 
\begin{eqnarray}
\nonumber \log[L_{\rm{1.8\,cm}} (\rm{mJy\,kpc}^2)] & = & -(1.81\pm0.29)\\
&&+(0.54\pm0.38)\log[L_{\rm{bol}} (\rm{L}_{\odot})],
\end{eqnarray}
\begin{eqnarray}
\nonumber \log[L_{\rm{1.8\,cm}} (\rm{mJy\,kpc}^2)] & = & -(1.23\pm0.65)\\
&&+(0.59\pm0.57)\log[L_{\rm{IR}} (\rm{L}_{\odot})],
\label{equ:corrIR}
\end{eqnarray}
with correlation coefficients of $r_{\rm{xy}}=0.77$ and 0.84, respectively. These regressions are fitted excluding data that are limits only. The correlation of bolometric luminosity with 1.8\,cm luminosity is similar to that found for 3.6\,cm data by Shirley et~al (2007) who found $ \log[L_{\rm{3.6\,cm}} (\rm{mJy\,kpc}^2)] = -(2.24\pm0.03)+(0.71\pm0.02)\log[L_{\rm{bol}} (\rm{L}_{\odot})]$. We note that there appears to be a trend in both the normalization and multiplicative factors in the correlations as a function of wavelength, although we note that the limited number of wavelengths sampled may make this trend misleading.
\begin{figure*}
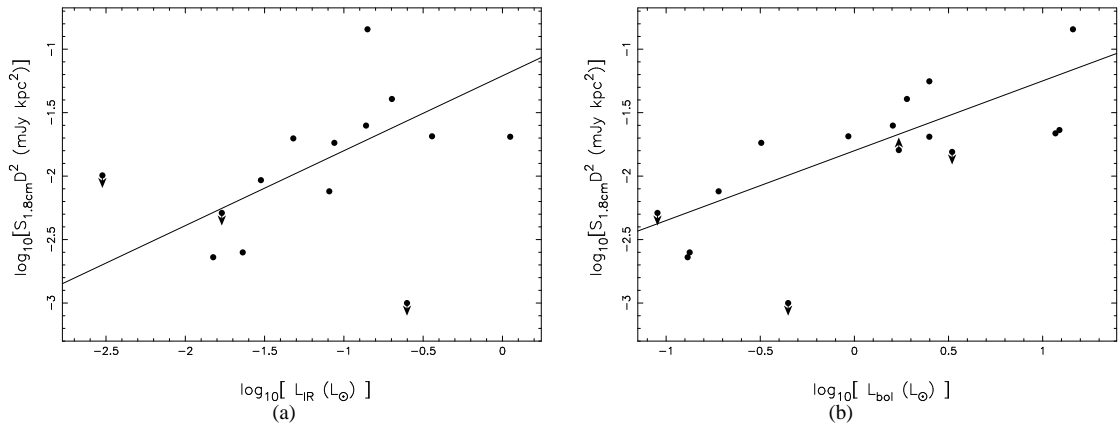

\centerline{\includegraphics[angle=-90,width=0.4\textwidth]{./corrIR.ps}\qquad\includegraphics[angle=-90,width=0.4\textwidth]{./corrBOL.ps}}
\centerline{(a)\hspace{0.4\textwidth}(b)}
\caption{Correlation of 1.8\,cm radio luminosity vs. (a) \emph{Spitzer} IR luminosity taken from Dunham et~al (2004); (b) bolometric luminosity for protostars, see Table~\ref{tab:phys} for references. Solid lines show linear regressions to the data. 
\label{fig:corr}}
\end{figure*}

The correlation between radio luminosity and $L_{\rm{IR}}$ provides a way of investigating the correlation with the internal luminosity of these sources. Dunham et~al. (2008) define VeLLO sources as those possessing an internal luminosity $L_{\rm{int}}\leq0.1$\,L$_{\odot}$ and use an approximately constant ratio of $L_{\rm{int}}/L_{\rm{IR}}=1.7$ to distinguish objects with $L_{\rm{IR}}\leq0.05$\,L$_{\odot}$ as VeLLOs. From Fig.~\ref{fig:corr}\,(a) we can see that 8 of our sample satisfy this criterion. Since $L_{\rm{int}}$ is not a directly observable quantity and may only be found through modelling the source, we can use the radio luminosity as a proxy for the internal luminosity of these objects by re-expressing the correlation in Equ.~\ref{equ:corrIR} as
\begin{equation}
L_{\rm{int}} ({\rm{L}}_{\odot}) \simeq 112 \times [L_{1.8\,\rm{cm}}(\rm{mJy\,kpc}^2)]^{1.7},
\end{equation}
or, following Dunham et~al. (2008) and normalizing to a distance of 200\,pc, we can further define the internal luminosity radio estimator as
\begin{equation}
L_{\rm{int}}^{\rm{rad}} ({\rm{L}}_{\odot}) = 1.3\times 10^{-5} [S_{1.8\,\rm{cm}} (\mu\rm{Jy})]^{1.7}.
\end{equation}

We do not consider this estimator to be as accurate as the 70\,$\mu$m flux estimator (Dunham et~al. 2008) as it relies on both the relationship of $L_{\rm{int}}$ to $L_{\rm{IR}}$ which is currently modelled as a simple ratio, and also because the correlation of $L_{1.8\rm{cm}}$ with $L_{\rm{IR}}$ is not as tightly defined as that of $F_{70}$ to $L_{\rm{int}}$. Two pieces of further work are required to improve the radio estimator of internal luminosity (1) a better constrained relationship between $L_{\rm{int}}$ and $L_{\rm{IR}}$, which uses additional data to investigate and fit the low luminosity excess of $L_{\rm{int}}/L_{\rm{IR}}$ above the constant ratio; and (2) more radio data to tighten the observed correlation between $L_{\rm{IR}}$ and $L_{\rm{rad}}$. However, it does provide a simple method to estimate the approximate evolutionary state of an object in the absence of complex models.

\subsection{The Non-Detections}
\label{sec:ND}

There are four protostellar (i.e. not starless) objects in our sample which are undetected by the AMI-LA at 1.8\,cm. The first of these is IRAS~04191. From our measured correlations both the $L_{\rm{IR}}$ and $L_{\rm{bol}}$ of this object should place it firmly above the detection threshold in our map. From the correlation of radio and bolometric luminosity we would predict a flux density of $S_{1.8\,\rm{cm}}\simeq518\,\mu$Jy. As IRAS~04191 lies approximately one-third of the way to the primary beam FWHM in our image we would expect to see IRAS~04191 detected at $\approx 25\,\sigma$ in Fig.~\ref{fig:figs1}\,(a). However, IRAS~04191 is an evolved Class~I object and if we make the assumption that the cm-wave emission we are seeing is a consequence of shock ionization due to a molecular outflow, we may explain the lack of radio emission from IRAS~04191 by the fact that Class~I objects can have outflows from an order of magnitude to several orders of magnitude weaker than those of Class~O sources (Bontemps et~al. 1996; Curtis et~al. 2010). The lack of cm-wave emission towards IRAS~04191 is in this case perhaps not surprising and may simply be due to a very weak outflow.

The second of our non-detections is L673-7-IRS. This object was identified as an embedded protostar by Dunham et~al. (2008) from the \emph{Spitzer} c2d data and has $L_{\rm{bol}}=0.09\pm0.03$. It was recently confirmed as a protostar through deep CO(2$\rightarrow$1) observations (Dunham et~al. 2010) which measured an outflow momentum flux of $F_{\rm{out}}>2.1\times10^{-6}$\,M$_{\odot}$\,yr$^{-1}$\,km\,s$^{-1}$ (or $F_{\rm{out}}>4.5\times10^{-6}$\,M$_{\odot}$\,yr$^{-1}$\,km\,s$^{-1}$ using the force per beam method (Fuller \& Ladd; Hatchell et~al. 2007). We can see from Fig.~\ref{fig:fcorr}(a) that with an outflow of such low momentum flux any associated radio emission may lie below our detection limit, possibly by more than an order of magnitude.

L1152-IRS is the third of our non-detections. The AMI-LA map of L1152-IRS shows some small structure at the pointing centre, coincident with the position of the sub-mm core. It is likely that the increased noise level in this dataset due to satellite interference has prevented us from making a significant detection of this object.

The final non-detection in our sample is L1148-IRS2, and indeed L1148-IRS1 may be considered a non-detection also. L1148-IRS2 is the very lowest $L_{\rm{IR}}$ object in our sample and was classified by Dunham et~al. (2008) as a Group 6 source under their classification scheme. Group 6 has the lowest probability of being a true protostar and it is possible that L1148-IRS2 is in fact an external galaxy. However, for a true source of such low predicted $L_{\rm{IR}}$ as L1148-IRS2, we can see from the radio-IR--luminosity correlation, Fig.~\ref{fig:corr}, that the AMI-LA observation would need to be at least an order of magnitude deeper in order to detect this object at 3\,$\sigma$. In the case of L1148-IRS1 it is possible that the \emph{Spitzer} spectrum is confused by an alternative physical process. It has been suggested (Hatchell \& Dunham 2009) that the Spitzer spectra of some low luminosity objects may confuse true embedded sources with H$_2$ shocks or PDR excitations (Smith et al. 2006). In this case such a hypothesis is supported by the lack of a compact sub-mm core in the SHARC-II 350\,$\mu$m observations of this field, which place an upper limit on the sub-mm flux density of $S_{350\,\mu\rm{m}} < 0.11$\,Jy (Wu et~al. 2007). Although a compact peak in the sub-mm does not conclusively indicate the presence of protostellar activity, it is difficult to reconcile such activity in the absence of one.
\begin{table}
\begin{center}
\caption{Detection Statistics. \label{tab:stats}}
\begin{tabular}{ccccc}
\hline \hline
Class & Total & Det. & Non-det. & Rate \\
&&&&(\%)\\
\hline
VeLLO & 8  & 5(2) & 3(3) & 63(100)\\
0     & 4* & 3    & 1*   & 75\\
I     & 8* & 6    & 2*   & 71\\
II    & 1* & 1*   & -    & 100\\
Starless & 3 & 0 & 3 & 0\\
\hline
\end{tabular}
\end{center}
\end{table}
\subsection{The Detections}

The detections and non-detections are summarized in Table~\ref{tab:stats}. The bracketed values indicate the number of VeLLOs within this bin that have not been confirmed independently of their \emph{Spitzer} identification and which may be confused by the physical mechanisms discussed in \S~\ref{sec:ND}. A starred value indicates a number which includes an object with a borderline class (e.g. L1152~SMM and Bern~48). The detection rate does not vary a great deal across the protostellar objects. A detection rate of zero towards starless objects confirms that the radio emission we see arises from a physical process other than a flattened dust tail due to vibrational emission. This highlights the ability of 16\,GHz radio observations to identify protostellar activity in cores which do not possess definitive sub-mm or infra-red measurements.

\section{Conclusions}
\label{sec:conc}

The rising radio spectrum of the cm-wave emission from protostellar objects makes higher radio frequencies more suitable for their detection than the longer wavelengths often employed. The low intensity of thermal dust emission at frequencies below $\approx 50$\,GHz greatly reduces the possibility of confusing the observed radio emission associated with molecular outflows or stellar winds, and therefore an indication of protostellar activity, with the cold dust tail from starless cores. The combination of these two effects makes 16\,GHz an effective frequency at which to distinguish protostellar objects and starless cores. This situation also extends to the very low luminosity objects, VeLLOs, and embedded objects since the dense cores surrounding these objects are transparent to the longer wavelengths of the radio emission. 

Two of the objects detected here, L723-IRS and L1165-IRS, were classified as poor candidates for embedded objects by Dunham et~al. (2008). However the cm-wave emission we detect towards these sources might suggest that some protostellar activity is present. As previously noted these objects may be external galaxies, or possibly later type stars which have been mis-identified in the MIR data, both of which could have radio counterparts. However, in both cases the morphology of the radio emission is not point-like, notably so in the case of L1165-IRS, which suggests that the radio emission is most probably not from a stellar wind in a later type star as this is generally well modelled by a spherical morphology and would appear point-like to the AMI-LA synthesized beam. If the radio emission arose from an external galaxy we would expect to see a steep spectrum synchrotron component and therefore also be able to identify the source in lower frequency radio data. More information is necessary to conclusively identify these objects as protostellar cores.

The radio emission from the sources detected here is generally $\ll 1$\,mJy and has a relatively compact morphology, although CB188~SMM-1 is possibly an exception to this situation. As such it should not provide a major source of confusion for spinning dust studies of the surrounding clouds that contain these cores, which are generally conducted on much larger angular scales.

The results presented here have shown that, in the absence of 70\,$\mu$m data, it is possible that the radio luminosity may be used as a proxy for the internal luminosity of embedded protostellar objects and to enable a larger sample of VeLLOs to be identified. However, at present the correlation presented here suffers from a lack of data and will require further observations to define the trend more completely.

\section{ACKNOWLEDGEMENTS} 

We thank the staff of the Lord's Bridge Observatory for their
invaluable assistance in the commissioning and operation of the
Arcminute Microkelvin Imager. We also thank the anonymous referee for their careful reading of this paper and useful comments. The AMI is supported by Cambridge University and the  STFC. CRG, TS, TF, MO and MLD   
acknowledge the support of PPARC/STFC studentships.

\bsp
\label{lastpage}


\begin{thebibliography}{}
\setlength{\labelwidth}{0pt}

\bibitem[\protect\citeauthoryear{Andr{\'e}, Motte, 
\& Bacmann}{1999}]{1999ApJ...513L..57A} Andr{\'e} P., Motte F., Bacmann A., 1999, ApJ, 513, L57 

\bibitem[\protect\citeauthoryear{Andr{\'e}, Ward-Thompson, 
\& Barsony}{2000}]{2000prpl.conf...59A} Andr{\'e} P., Ward-Thompson D., Barsony M., 2000, prpl.conf, 59 

\bibitem[\protect\citeauthoryear{Andr{\'e}, Montmerle, 
\& Feigelson}{1987}]{1987AJ.....93.1182A} Andr{\'e} P., Montmerle T., Feigelson E.~D., 1987, AJ, 93, 1182 

\bibitem[\protect\citeauthoryear{Anglada}{1996}]{1996ASPC...93....3A} 
Anglada G., 1996, ASPC, 93, 3 

\bibitem[\protect\citeauthoryear{Anglada}{1995}]{1995RMxAC...1...67A} 
Anglada G., 1995, RMxAC, 1, 67 

\bibitem[\protect\citeauthoryear{Anglada et 
al.}{1991}]{1991ApJ...376..615A} Anglada G., Estalella R., Rodr{\'i}guez L.~F., 
Torrelles J.~M., Lopez R., Canto J., 1991, ApJ, 376, 615 

\bibitem[\protect\citeauthoryear{Armstrong 
\& Winnewisser}{1989}]{1989A&A...210..373A} Armstrong J.~T., Winnewisser G., 1989, A\&A, 210, 373 


\bibitem[\protect\citeauthoryear{Bieging, Cohen, 
\& Schwartz}{1984}]{1984ApJ...282..699B} Bieging J.~H., Cohen M., Schwartz P.~R., 1984, ApJ, 282, 699 

\bibitem[\protect\citeauthoryear{Bontemps et 
al.}{1996}]{1996A&A...311..858B} Bontemps S., Andr{\'e} P., Terebey S., Cabrit S., 1996, A\&A, 311, 858 

\bibitem[\protect\citeauthoryear{Bourke et al.}{2006}]{2006ApJ...649L..37B} 
Bourke T.~L., et al., 2006, ApJ, 649, L37 


\bibitem[\protect\citeauthoryear{Bourke et al.}{2005}]{2005ApJ...633L.129B} 
Bourke T.~L., Crapsi A., Myers P.~C., Evans N.~J., II, Wilner D.~J., Huard 
T.~L., J{\o}rgensen J.~K., Young C.~H., 2005, ApJ, 633, L129 


\bibitem[\protect\citeauthoryear{Carrasco-Gonz{\'a}lez et 
al.}{2008}]{2008ApJ...676.1073C} Carrasco-Gonz{\'a}lez C., Anglada G., 
Rodr{\'{\i}}guez L.~F., Torrelles J.~M., Osorio M., Girart J.~M., 2008, 
ApJ, 676, 1073 


\bibitem[\protect\citeauthoryear{Casassus et 
al.}{2006}]{2006ApJ...639..951C} Casassus S., Cabrera G.~F., F{\"o}rster 
F., Pearson T.~J., Readhead A.~C.~S., Dickinson C., 2006, ApJ, 639, 951 


\bibitem[\protect\citeauthoryear{Cassen 
\& Moosman}{1981}]{1981Icar...48..353C} Cassen P., Moosman A., 1981, Icar, 48, 353 


\bibitem[\protect\citeauthoryear{Churchwell}{1990}]{1990A&ARv...2...79C} Churchwell E., 1990, A\&ARv, 2, 79 

\bibitem[\protect\citeauthoryear{Connelley, Reipurth, 
\& Tokunaga}{2008}]{2008AJ....135.2496C} Connelley M.~S., Reipurth B., Tokunaga A.~T., 2008, AJ, 135, 2496 

\bibitem[\protect\citeauthoryear{Crapsi et 
al.}{2005}]{2005A&A...439.1023C} Crapsi A., et al., 2005, A\&A, 439, 1023 


\bibitem[\protect\citeauthoryear{Curiel et 
al.}{1989}]{1989ApL&C..27..299C} Curiel S., Rodr{\'i}guez L.~F., Bohigas J., Roth M., Canto J., Torrelles J.~M., 1989, ApL\&C, 27, 299 


\bibitem[\protect\citeauthoryear{Curiel, Canto, 
\& Rodr{\'i}guez}{1987}]{1987RMxAA..14..595C} Curiel S., Canto J., Rodr{\'i}guez L.~F., 1987, RMxAA, 14, 595 

\bibitem[\protect\citeauthoryear{Curtis et al.}{2010}]{2010arXiv1006.3218C} 
Curtis E.~I., Richer J.~S., Swift J.~J., Williams J.~P., 2010, arXiv, 
arXiv:1006.3218 


\bibitem[\protect\citeauthoryear{de Zotti et 
al.}{2005}]{2005A&A...431..893D} de Zotti G., Ricci R., Mesa D., Silva L., Mazzotta P., Toffolatti L., Gonz{\'a}lez-Nuevo J., 2005, A\&A, 431, 893 

\bibitem[\protect\citeauthoryear{Draine 
\& Lazarian}{1998}]{1998ApJ...508..157D} Draine B.~T., Lazarian A., 1998, ApJ, 508, 157 

\bibitem[\protect\citeauthoryear{Dunham et al.}{2010}]{2010arXiv1008.1049D} 
Dunham M.~M., Evans N.~J., II, Bourke T.~L., Myers P.~C., Huard T.~L., 
Stutz A.~M., 2010, arXiv, arXiv:1008.1049 

\bibitem[\protect\citeauthoryear{Dunham et al.}{2008}]{2008ApJS..179..249D} 
Dunham M.~M., Crapsi A., Evans N.~J., II, Bourke T.~L., Huard T.~L., Myers 
P.~C., Kauffmann J., 2008, ApJS, 179, 249 

\bibitem[\protect\citeauthoryear{Dunham et al.}{2006}]{2006ApJ...651..945D} 
Dunham M.~M., et al., 2006, ApJ, 651, 945 

\bibitem[\protect\citeauthoryear{Enoch et al.}{2007}]{2007ApJ...666..982E} 
Enoch M.~L., Glenn J., Evans N.~J., II, Sargent A.~I., Young K.~E., Huard 
T.~L., 2007, ApJ, 666, 982 

\bibitem[\protect\citeauthoryear{Evans et al.}{2009}]{2009ApJS..181..321E} 
Evans N.~J., et al., 2009, ApJS, 181, 321 

\bibitem[\protect\citeauthoryear{Evans et al.}{2003}]{2003PASP..115..965E} 
Evans N.~J., II, et al., 2003, PASP, 115, 965 

\bibitem[\protect\citeauthoryear{Fuller 
\& Ladd}{2002}]{2002ApJ...573..699F} Fuller G.~A., Ladd E.~F., 2002, ApJ, 573, 699 

\bibitem[\protect\citeauthoryear{Furuya et al.}{2003}]{2003ApJS..144...71F} 
Furuya R.~S., Kitamura Y., Wootten A., Claussen M.~J., Kawabe R., 2003, 
ApJS, 144, 71 

\bibitem[\protect\citeauthoryear{Harvey et al.}{2002}]{2002AJ....123.3325H} 
Harvey D.~W.~A., Wilner D.~J., Di Francesco J., Lee C.~W., Myers P.~C., 
Williams J.~P., 2002, AJ, 123, 3325 

\bibitem[\protect\citeauthoryear{Hatchell 
\& Dunham}{2009}]{2009A&A...502..139H} Hatchell J., Dunham M.~M., 2009, A\&A, 502, 139 

\bibitem[\protect\citeauthoryear{Hatchell, Fuller, 
\& Richer}{2007}]{2007A&A...472..187H} Hatchell J., Fuller G.~A., Richer J.~S., 2007, A\&A, 472, 187 

\bibitem[\protect\citeauthoryear{Hayashi et 
al.}{1994}]{1994ApJ...426..234H} Hayashi M., Hasegawa T., Ohashi N., Sunada 
K., 1994, ApJ, 426, 234 


\bibitem[\protect\citeauthoryear{Hurley-Walker et 
al.}{2009}]{2009MNRAS.398..249H} AMI Consortium: Hurley-Walker N., et al., 2009, MNRAS, 
398, 249 

\bibitem[\protect\citeauthoryear{Kauffmann et 
al.}{2008}]{2008A&A...487..993K} Kauffmann J., Bertoldi F., Bourke T.~L., Evans N.~J., II, Lee C.~W., 2008, A\&A, 487, 993 

\bibitem[\protect\citeauthoryear{Kauffmann et 
al.}{2005}]{2005AN....326..878K} Kauffmann J., Bertoldi F., Evans N.~J., 
II, the C2D Collaboration, 2005, AN, 326, 878 

\bibitem[\protect\citeauthoryear{Kenyon et al.}{1990}]{1990AJ.....99..869K} 
Kenyon S.~J., Hartmann L.~W., Strom K.~M., Strom S.~E., 1990, AJ, 99, 869 

\bibitem[\protect\citeauthoryear{Kenyon 
\& Hartmann}{1995}]{1995ApJS..101..117K} Kenyon S.~J., Hartmann L., 1995, ApJS, 101, 117 

\bibitem[\protect\citeauthoryear{Kirk, Ward-Thompson, 
\& Andr{\'e}}{2007}]{2007MNRAS.375..843K} Kirk J.~M., Ward-Thompson D., Andr{\'e} P., 2007, MNRAS, 375, 843 

\bibitem[\protect\citeauthoryear{Konigl}{1982}]{1982ApJ...261..115K} Konigl 
A., 1982, ApJ, 261, 115 

\bibitem[\protect\citeauthoryear{Launhardt 
\& Henning}{1997}]{1997A&A...326..329L} Launhardt R., Henning T., 1997, A\&A, 326, 329 

\bibitem[\protect\citeauthoryear{Lee 
\& Ho}{2005}]{2005ApJ...632..964L} Lee C.-F., Ho P.~T.~P., 2005, ApJ, 632, 964 

\bibitem[\protect\citeauthoryear{Lee, Myers, 
\& Plume}{2004}]{2004ApJS..153..523L} Lee C.~W., Myers P.~C., Plume R., 2004, ApJS, 153, 523 

\bibitem[\protect\citeauthoryear{Myers et al.}{1988}]{1988ApJ...324..907M} 
Myers P.~C., Heyer M., Snell R.~L., Goldsmith P.~F., 1988, ApJ, 324, 907 

\bibitem[\protect\citeauthoryear{Ohashi et al.}{1996}]{1996ApJ...466..317O} 
Ohashi N., Hayashi M., Kawabe R., Ishiguro M., 1996, ApJ, 466, 317 


\bibitem[\protect\citeauthoryear{Panagia 
\& Felli}{1975}]{1975A&A....39....1P} Panagia N., Felli M., 1975, A\&A, 39, 1 


\bibitem[\protect\citeauthoryear{Patnaik et 
al.}{1992}]{1992MNRAS.254..655P} Patnaik A.~R., Browne I.~W.~A., Wilkinson 
P.~N., Wrobel J.~M., 1992, MNRAS, 254, 655 


\bibitem[\protect\citeauthoryear{Pravdo et al.}{1985}]{1985ApJ...293L..35P} 
Pravdo S.~H., Rodr{\'i}guez L.~F., Curiel S., Canto J., Torrelles J.~M., Becker 
R.~H., Sellgren K., 1985, ApJ, 293, L35

\bibitem[\protect\citeauthoryear{Reynolds}{1986}]{1986ApJ...304..713R} 
Reynolds S.~P., 1986, ApJ, 304, 713 


\bibitem[\protect\citeauthoryear{Rodr{\'{\i}}guez 
\& Reipurth}{1998}]{1998RMxAA..34...13R} Rodr{\'{\i}}guez L.~F., Reipurth B., 1998, RMxAA, 34, 13 


\bibitem[\protect\citeauthoryear{Rodr{\'{\i}}guez 
\& Reipurth}{1996}]{1996RMxAA..32...27R} Rodr{\'{\i}}guez L.~F., Reipurth B., 1996, RMxAA, 32, 27 


\bibitem[\protect\citeauthoryear{Rodr{\'i}guez}{1995}]{1995RMxAC...1....1R} 
Rodr{\'i}guez L.~F., 1995, RMxAC, 1, 1 


\bibitem[\protect\citeauthoryear{Rodr{\'i}guez et 
al.}{1994}]{1994ApJ...427L.103R} Rodr{\'i}guez L.~F., Canto J., Torrelles 
J.~M., Gomez J.~F., Anglada G., Ho P.~T.~P., 1994, ApJ, 427, L103 


\bibitem[\protect\citeauthoryear{Rodr{\'i}guez et 
al.}{1989}]{1989ApJ...347..461R} Rodr{\'i}guez L.~F., Myers P.~C., 
Cruz-Gonzalez I., Terebey S., 1989, ApJ, 347, 461 

\bibitem[\protect\citeauthoryear{Rodr{\'i}guez 
\& Canto}{1983}]{1983RMxAA...8..163R} Rodr{\'i}guez L.~F., Canto J., 1983, RMxAA, 8, 163 


\bibitem[\protect\citeauthoryear{Scaife et al.}{2010}]{2010MNRAS.403L..46S} 
AMI Consortium: Scaife A.~M.~M., et al., 2010, MNRAS, 403, L46 


\bibitem[\protect\citeauthoryear{Scaife et al.}{2009}]{2009MNRAS.400.1394S} 
AMI Consortium: Scaife A.~M.~M., et al., 2009a, MNRAS, 400, 1394 


\bibitem[\protect\citeauthoryear{Scaife et 
al.}{2009}]{2009MNRAS.394L..46A} AMI Consortium: Scaife A.~M.~M., et al., 2009b, MNRAS, 394, 
L46 

\bibitem[\protect\citeauthoryear{Scaife et al.}{2008}]{2008MNRAS.385..809S} 
AMI Consortium: Scaife A.~M.~M., et al., 2008, MNRAS, 385, 809 


\bibitem[\protect\citeauthoryear{Shang}{2004}]{2004IAUS..221..351S} Shang 
H., 2004, IAUS, 221, 351 

\bibitem[\protect\citeauthoryear{Shinnaga et 
al.}{2009}]{2009ApJ...706L.226S} Shinnaga H., Phillips T.~G., Furuya R.~S., 
Kitamura Y., 2009, ApJ, 706, L226 


\bibitem[\protect\citeauthoryear{Shirley et 
al.}{2007}]{2007ApJ...667..329S} Shirley Y.~L., Claussen M.~J., Bourke 
T.~L., Young C.~H., Blake G.~A., 2007, ApJ, 667, 329 


\bibitem[\protect\citeauthoryear{Shu}{1977}]{1977ApJ...214..488S} Shu 
F.~H., 1977, ApJ, 214, 488 

\bibitem[\protect\citeauthoryear{Simon et al.}{1983}]{1983ApJ...266..623S} 
Simon M., Felli M., Massi M., Cassar L., Fischer J., 1983, ApJ, 266, 623 

\bibitem[\protect\citeauthoryear{Smith et al.}{2006}]{2006ApJ...645.1264S} 
Smith H.~A., Hora J.~L., Marengo M., Pipher J.~L., 2006, ApJ, 645, 1264 

\bibitem[\protect\citeauthoryear{Stamatellos et 
al.}{2007}]{2007A&A...462..677S} Stamatellos D., Ward-Thompson D., Whitworth A.~P., Bontemps S., 2007, A\&A, 462, 677 

\bibitem[\protect\citeauthoryear{Terebey et 
al.}{2009}]{2009ApJ...696.1918T} Terebey S., et al., 2009, ApJ, 696, 1918 

\bibitem[\protect\citeauthoryear{Umemoto et 
al.}{1991}]{1991ApJ...377..510U} Umemoto T., Hirano N., Kameya O., Fukui 
Y., Kuno N., Takakubo K., 1991, ApJ, 377, 510 


\bibitem[\protect\citeauthoryear{Visser, Richer, 
\& Chandler}{2002}]{2002AJ....124.2756V} Visser A.~E., Richer J.~S., Chandler C.~J., 2002, AJ, 124, 2756 

\bibitem[\protect\citeauthoryear{Waldram et 
al.}{2010}]{2010MNRAS.404.1005W} Waldram E.~M., Pooley G.~G., Davies M.~L., 
Grainge K.~J.~B., Scott P.~F., 2010, MNRAS, 404, 1005 

\bibitem[\protect\citeauthoryear{Winkler 
\& Newman}{1980}]{1980ApJ...236..201W} Winkler K.-H.~A., Newman M.~J., 1980, ApJ, 236, 201 

\bibitem[\protect\citeauthoryear{Wright 
\& Barlow}{1975}]{1975MNRAS.170...41W} Wright A.~E., Barlow M.~J., 1975, MNRAS, 170, 41 


\bibitem[\protect\citeauthoryear{Wu et al.}{2007}]{2007AJ....133.1560W} Wu 
J., Dunham M.~M., Evans N.~J., II, Bourke T.~L., Young C.~H., 2007, AJ, 
133, 1560 


\bibitem[\protect\citeauthoryear{Young et al.}{2009}]{2009ApJ...702..340Y} 
Young C.~H., et al., 2009, ApJ, 702, 340 


\bibitem[\protect\citeauthoryear{Young et al.}{2006}]{2006AJ....132.1998Y} 
Young C.~H., Bourke T.~L., Young K.~E., Evans N.~J., II, J{\o}rgensen 
J.~K., Shirley Y.~L., van Dishoeck E.~F., Hogerheijde M., 2006, AJ, 132, 
1998 


\bibitem[\protect\citeauthoryear{Young 
\& Evans}{2005}]{2005ApJ...627..293Y} Young C.~H., Evans N.~J., II, 2005, ApJ, 627, 293 

\bibitem[\protect\citeauthoryear{Young et al.}{2004}]{2004ApJS..154..396Y} 
Young C.~H., et al., 2004, ApJS, 154, 396 

\bibitem[\protect\citeauthoryear{Yun 
\& Clemens}{1994}]{1994ApJS...92..145Y} Yun J.~L., Clemens D.~P., 1994, ApJS, 92, 145 

\bibitem[\protect\citeauthoryear{Zwart et al.}{2008}]{2008MNRAS.391.1545Z} 
AMI Consortium: Zwart J.~T.~L., et al., 2008, MNRAS, 391, 1545 






\end{thebibliography}
\end{document}